\documentstyle[12pt,graphicx,amsmath,tabularx,enumitem,multirow]{article}
\newlength{\pubnumber} 

\catcode`\@=11
\@addtoreset{equation}{section}

\def\section{\@startsection{section}{1}{\z@}{3.5ex plus 1ex minus .2ex}
 {2.3ex plus .2ex}{\large\bf}}
\def\subsection{\@startsection{subsection}{2}{\z@}{2.3ex plus .2ex}
 {2.3ex plus .2ex}{\bf}}

\newcommand{\ba}{\begin{eqnarray}}
\newcommand{\ea}{\end{eqnarray}}
\begin{document}

\begin{titlepage}
\samepage{
\setcounter{page}{1}
\rightline{LTH--1044}

\vfill
\begin{center}
 {\Large \bf
Non--Tachyonic Semi--Realistic \\Non--Supersymmetric Heterotic String Vacua}
\vspace{1cm}
\vfill {\large
Johar M. Ashfaque,\\
\vspace{.1in}

 Panos Athanasopoulos, Alon E. Faraggi
and Hasan Sonmez
}\\
\vspace{1cm}
{\it Dept.\ of Mathematical Sciences,
             University of Liverpool,
         Liverpool L69 7ZL, UK\footnote{Emails: jauhar@liv.ac.uk,
panos@liv.ac.uk, Alon.Faraggi@liv.ac.uk, Hasan.Sonmez@liv.ac.uk}\\}
\vspace{.025in}
\end{center}
\vfill
\begin{abstract}

\noindent

The heterotic--string models in the free fermionic formulation gave rise
to some of the most realistic string models to date, which possess
$N=1$ spacetime supersymmetry. Lack of evidence for supersymmetry
at the LHC instigated recent interest in non--supersymmetric
heterotic--string vacua. We explore what may be learned
in this context from the quasi--realistic free fermionic
models. We show that constructions with a low number of
families give rise to proliferation of a priori tachyon producing sectors,
compared to the non--realistic examples, which typically
may contain only one such sector. The reason being that
in the realistic cases the internal six dimensional
space is fragmented into smaller units.
We present one example of a quasi--realistic, non--supersymmetric,
non--tachyonic, heterotic--string vacuum and compare
the structure of its massless spectrum to the corresponding
supersymmetric vacuum. While in some sectors supersymmetry
is broken explicitly, {\it i.e.}~the bosonic and fermionic
sectors produce massless and massive states,
other sectors, and in particular those leading to the chiral
families, continue to exhibit fermi--bose degeneracy.
In these sectors the massless spectrum, as compared
to the supersymmetric cases, will only differ in some
local or global $U(1)$ charges.
We discuss the conditions for obtaining $n_b=n_f$ at the
massless level in these models.
Our example model contains an anomalous $U(1)$
symmetry, which generates a tadpole diagram at
one loop--order in string perturbation theory.
We speculate that this tadpole diagram
may cancel the corresponding diagram generated
by the one--loop non--vanishing
vacuum energy and that in this respect
the supersymmetric and non--supersymmetric
vacua should be regarded on equal footing. Finally we
discuss vacua that contain two supersymmetry generating
sectors.

\end{abstract}
\smallskip}
\end{titlepage}

\setcounter{footnote}{0}

\def\beq{\begin{equation}}
\def\eeq{\end{equation}}
\def\beqn{\begin{eqnarray}}
\def\eeqn{\end{eqnarray}}

\def\no{\noindent }
\def\nolabel{\nonumber }
\def\ie{{\it i.e.}}
\def\eg{{\it e.g.}}
\def\half{{\textstyle{1\over 2}}}
\def\third{{\textstyle {1\over3}}}
\def\quarter{{\textstyle {1\over4}}}
\def\sixth{{\textstyle {1\over6}}}
\def\m{{\tt -}}
\def\p{{\tt +}}

\def\Tr{{\rm Tr}\, }
\def\tr{{\rm tr}\, }

\def\slash#1{#1\hskip-6pt/\hskip6pt}
\def\slk{\slash{k}}
\def\GeV{\,{\rm GeV}}
\def\TeV{\,{\rm TeV}}
\def\y{\,{\rm y}}
\def\SM{Standard--Model }
\def\SUSY{supersymmetry }
\def\SSSM{supersymmetric standard model}
\def\vev#1{\left\langle #1\right\rangle}
\def\l{\langle}
\def\r{\rangle}
\def\o#1{\frac{1}{#1}}

\def\Htw{{\tilde H}}
\def\chibar{{\overline{\chi}}}
\def\qbar{{\overline{q}}}
\def\ibar{{\overline{\imath}}}
\def\jbar{{\overline{\jmath}}}
\def\Hbar{{\overline{H}}}
\def\Qbar{{\overline{Q}}}
\def\abar{{\overline{a}}}
\def\alphabar{{\overline{\alpha}}}
\def\betabar{{\overline{\beta}}}
\def\tautwo{{ \tau_2 }}
\def\thetatwo{{ \vartheta_2 }}
\def\thetathree{{ \vartheta_3 }}
\def\thetafour{{ \vartheta_4 }}
\def\ttwo{{\vartheta_2}}
\def\tthree{{\vartheta_3}}
\def\tfour{{\vartheta_4}}
\def\ti{{\vartheta_i}}
\def\tj{{\vartheta_j}}
\def\tk{{\vartheta_k}}
\def\calF{{\cal F}}
\def\smallmatrix#1#2#3#4{{ {{#1}~{#2}\choose{#3}~{#4}} }}
\def\ab{{\alpha\beta}}
\def\Minv{{ (M^{-1}_\ab)_{ij} }}
\def\bone{{\bf 1}}
\def\ii{{(i)}}
\def\V{{\bf V}}
\def\N{{\bf N}}

\def\b{{\bf b}}
\def\S{{\bf S}}
\def\X{{\bf X}}
\def\I{{\bf I}}
\def\mb{{\mathbf b}}
\def\mS{{\mathbf S}}
\def\mX{{\mathbf X}}
\def\mI{{\mathbf I}}
\def\balpha{{\mathbf \alpha}}
\def\bbeta{{\mathbf \beta}}
\def\bgamma{{\mathbf \gamma}}
\def\bxi{{\mathbf \xi}}

\def\t#1#2{{ \Theta\left\lbrack \matrix{ {#1}\cr {#2}\cr }\right\rbrack }}
\def\C#1#2{{ C\left\lbrack \matrix{ {#1}\cr {#2}\cr }\right\rbrack }}
\def\tp#1#2{{ \Theta'\left\lbrack \matrix{ {#1}\cr {#2}\cr }\right\rbrack }}
\def\tpp#1#2{{ \Theta''\left\lbrack \matrix{ {#1}\cr {#2}\cr }\right\rbrack }}
\def\l{\langle}
\def\r{\rangle}
\newcommand{\cc}[2]{c{#1\atopwithdelims()#2}}
\newcommand{\nn}{\nonumber}


\def\inbar{\,\vrule height1.5ex width.4pt depth0pt}

\def\IC{\relax\hbox{$\inbar\kern-.3em{\rm C}$}}
\def\IQ{\relax\hbox{$\inbar\kern-.3em{\rm Q}$}}
\def\IR{\relax{\rm I\kern-.18em R}}
 \font\cmss=cmss10 \font\cmsss=cmss10 at 7pt
\def\IZ{\relax\ifmmode\mathchoice
 {\hbox{\cmss Z\kern-.4em Z}}{\hbox{\cmss Z\kern-.4em Z}}
 {\lower.9pt\hbox{\cmsss Z\kern-.4em Z}}
 {\lower1.2pt\hbox{\cmsss Z\kern-.4em Z}}\else{\cmss Z\kern-.4em Z}\fi}

\def\AEF{A.E. Faraggi}
\def\JHEP#1#2#3{{\it JHEP}\/ {\bf #1} (#2) #3}
\def\NPB#1#2#3{{\it Nucl.\ Phys.}\/ {\bf B#1} (#2) #3}
\def\PLB#1#2#3{{\it Phys.\ Lett.}\/ {\bf B#1} (#2) #3}
\def\PRD#1#2#3{{\it Phys.\ Rev.}\/ {\bf D#1} (#2) #3}
\def\PRL#1#2#3{{\it Phys.\ Rev.\ Lett.}\/ {\bf #1} (#2) #3}
\def\PRT#1#2#3{{\it Phys.\ Rep.}\/ {\bf#1} (#2) #3}
\def\MODA#1#2#3{{\it Mod.\ Phys.\ Lett.}\/ {\bf A#1} (#2) #3}
\def\IJMP#1#2#3{{\it Int.\ J.\ Mod.\ Phys.}\/ {\bf A#1} (#2) #3}
\def\nuvc#1#2#3{{\it Nuovo Cimento}\/ {\bf #1A} (#2) #3}
\def\RPP#1#2#3{{\it Rept.\ Prog.\ Phys.}\/ {\bf #1} (#2) #3}
\def\EJP#1#2#3{{\it Eur.\ Phys.\ Jour.}\/ {\bf C#1} (#2) #3}
\def\etal{{\it et al\/}}

\hyphenation{su-per-sym-met-ric non-su-per-sym-met-ric}
\hyphenation{space-time-super-sym-met-ric}
\hyphenation{mod-u-lar mod-u-lar--in-var-i-ant}


\setcounter{footnote}{0}
\section{Introduction}

The discovery of the agent of electroweak symmetry breaking
at the LHC \cite{higgsdiscovery} is a pivotal moment in particle physics.
While confirmation of this agent as the Standard Model electroweak doublet
representation will require experimental scrutiny in the decades to come,
the data to date seems to vindicate this possibility.
Substantiation of this interpretation of the data
will reinforce the view that the electroweak
symmetry breaking mechanism is intrinsically
perturbative, and that the SM provides a viable perturbative
parameterisation up to the Planck scale. Moreover, the large
scale unification scenario is further motivated by the embedding
of the SM matter states in the chiral $SO(10)$ representation; by the
logarithmic evolution of the SM parameters;
by proton longevity; and by the suppression of left--handed neutrino masses.
Gaining further insight
into the fundamental origins of the SM parameters
can then only be obtained by incorporating
gravity into the picture.

String theory provides
the most developed contemporary approach to
study how the Standard Model parameters
may arise from a unified theory of the
gauge and gravitational interactions.
For this purpose several models that
reproduce the spectrum of the Minimal
Supersymmetric Standard Model have been produced
\cite{fny,stringmssm}.
Amongst them the free fermionic models
\cite{fny, fsu5, slm, alr, lrs, su421, fknr, acfkr, su62, frs}
are the most studied examples.
The heterotic string in particular provides
a compelling framework to study the
gauge -- gravity synthesis in the
large scale unification scenario,
as it reproduces the embedding of the SM chiral
spectrum in spinorial $SO(10)$ representations.

The majority of semi--realistic heterotic string models constructed
to date possess $N=1$ spacetime  supersymmetry,
while non--supersymmetric vacua were investigated sporadically
\cite{aggmv, ft2, oldnonsusy, ft1}.
In the absence
of evidence of supersymmetry at the LHC recent interest
in non--supersymmetric heterotic string vacua has emerged
\cite{stefan, fkp, abeldienes, florakis, lukas}.
It is therefore prudent to examine what may be learned
in that context from the quasi--realistic free fermionic models.
In this paper this question is considered. We
discuss the different avenues that may be used
to break supersymmetry directly at the string
scale and how they compare with the recent analysis \cite{fkp}.

Our paper is organised as follows: in section \ref{review},
we review the structure of the phenomenological free
fermionic heterotic string models. In section \ref{tachfree}, we discuss the
phases that break supersymmetry in the string models
and the different patterns that they induce.
Further discussion on the existence of sectors
producing tachyons in these models and the relation of the abundance of these
sectors with the number of families is given. Moreover, in section
\ref{explicitmodel} a non--supersymmetric tachyon free model is presented and
its relation to the supersymmetric counterpart is inferred.
In section \ref{split} we discuss the construction
of string vacua with split supersymmetry, in which
supersymmetry is produced by two sectors.
Section \ref{conclusions} contains our conclusions.

\section{Phenomenological free fermionic models}\label{review}

In this section, we review the structure of the phenomenological free fermionic
models. It should be stressed that these free fermionic models correspond to
$Z_2\times Z_2$ toroidal orbifolds and their phenomenological characteristics
are deeply rooted in the structure of the $Z_2\times Z_2$ orbifolds.
In this respect, the free fermionic formalism merely provides an accessible set
of tools to extract the spectra of the string vacua and their properties.
Furthermore, the free fermionic machinery
extends to the massive string spectrum via the analysis of the relevant
partition function. This provides important insight into the symmetries that
underly the string landscape and eventually may prove instrumental in
understanding how the string
vacuum is selected. However, one should not tie the cart before the horse. The
fermionic and bosonic representations only provide complementary tools
that are formally identical in two dimensions. The physically relevant
properties of these free fermionic models are due to their underlying
$Z_2\times Z_2$ orbifold structure.

In the free fermionic formulation of the heterotic string in four dimensions,
all the extra degrees of freedom needed to cancel the conformal anomaly are
represented as free fermions propagating on
the string worldsheet. It is important to note that
the two dimensional fermions are free only at a special point in the moduli
space \cite{ginsparg}. However, the models can be deformed away from that point
by incorporating the moduli as worldsheet Thirring interactions
\cite{changkumar}. Since the twisted matter spectrum of the $Z_2\times Z_2$
orbifolds,
which gives rise to the Standard Model matter states, is independent of the
moduli, working at the free fermionic point is just a convenient choice. In the
light--cone gauge the supersymmetric
left--moving sector includes the two transverse spacetime fermionic
coordinates $\psi^\mu$ and 18 internal worldsheet real fermions $\chi^I$,
whereas the right--moving bosonic sector contains 44 real worldsheet
fermions $\phi^a$. The worldsheet supersymmetry is realised non--linearly
in the left--moving sector and the worldsheet supercurrent is given by
\beq
T_F = \psi^\mu\partial X_\mu +f_{IJK}\chi^I\chi^J\chi^K~,
\label{supercurrent}
\eeq
where $f_{IJK}$ are the structure constants of the 18 dimensional semi--simple
Lie group. The 18 left--moving worldsheet fermions $\chi^I$
transform in the adjoint representation of the Lie group, which in the case of
the
fermionic $Z_2\times Z_2$ orbifolds with $N=1$ SUSY is $SU(2)^6$. Such models
provide our starting point and we will discuss in later sections how
supersymmetry is broken. The $\chi^I$ therefore
transform in the adjoint representation of $SU(2)^6$, and are denoted by
$\chi^I, y^I,\omega^I$ with $I=1,\cdots, 6$. Under parallel transport
around a non--contractible loop of the one--loop vacuum to vacuum amplitude the
worldsheet fermions pick up a phase \beq
f~\rightarrow~-{\rm e}^{i\pi\alpha(f)}f~,
\label{fermionphases}
\eeq
with $\alpha(f)~\in~(-1, +1]$. The phases for all worldsheet fermions
constitute
the spin structure of the models and are given in the form of 64 dimensional
boundary condition basis vectors. The partition function
\beq
Z{(\tau)}~=~ \sum_{\alpha, \beta \in \Xi}
 \cc{\alpha}{\beta} {\rm Tr}
{\alpha\atopwithdelims()\beta},
\eeq
is a sum over all spin structures, where $\cc{\alpha}{\beta}$ are
Generalised GSO (GGSO) projection coefficients and ${\rm Tr}
{\alpha\atopwithdelims()\beta}\equiv \rm{Tr}(e^{i\pi\beta F_\alpha} e^{i\pi\tau
H_\alpha})$ with $H_\alpha$ being the hamiltonian, is the trace over the mode
excitations
of the worldsheet fields in the sector $\alpha$, subject to the GSO
projections induced by the sector $\beta$.
Requiring invariance under modular transformations results in a
set of constraints on the allowed spin structures and the GGSO
projection coefficients.
The Hilbert space of a given sector $\alpha$ in the finite Abelian additive
group $\alpha\in \Xi={\sum_k}n_i {b}_i$,
where $n_i=0,\cdots,{{g_{z_i}}-1}$, is obtained by
acting on the vacuum of the sector $\alpha$ with bosonic, as well as
fermionic oscillators with frequencies $\nu_f$, $\nu_{f^*}$,
and subsequently imposing the GGSO projections
\begin{equation}
\left\{e^{i\pi({b_i}F_\alpha)}-
{\delta_\alpha}
c^*\left(
\begin{matrix}
\alpha\cr
b_i\cr
\end{matrix}
\right)
\right\}\vert{s}\rangle_\alpha=0
\label{gsoprojections}
\end{equation}
with
\begin{equation}
(b_i{F_\alpha})\equiv\{\sum_{real+complex\atop{left}}-
\sum_{real+complex\atop{right}}\}(b_i(f)F_\alpha(f)),
\label{lorentzproduct}
\end{equation}
where $\delta_\alpha$ is the spacetime spin statistics index and
$F_\alpha(f)$ is a fermion number operator, counting each mode of
$f$ once (and if $f$ is complex, $f^*$ minus once).
For Ramond fermions with $\alpha(f)=1$ the vacuum is a doubly
degenerate spinor $\vert\pm\rangle$, annihilated by the zero modes
$f_0$ and $f^*_0$, and with fermion numbers $F(f)=0,~-1$.
The physical states in the string spectrum satisfy the
level matching condition
\beq
M_L^2=-{1\over 2}+{{{\alpha_L}\cdot{\alpha_L}}\over 8}+N_L=-1+
{{{\alpha_R}\cdot{\alpha_R}}\over 8}+N_R=M_R^2
\label{virasorocond}
\eeq
where
$\alpha=(\alpha_L;\alpha_R)\in\Xi$ is a sector in the additive group, and
\beq
N_L=\sum_f ({\nu_L}) ;\hskip 3cm N_R=\sum_f ({\nu_R});
\label{nlnr}
\eeq
\beq
\nu_f={{1+\alpha(f)}\over 2} ;\hskip 3cm
{\nu_{f^*}}={{1-\alpha(f)}\over 2}.
\label{nulnur}
\eeq
The $U(1)$ charges with respect to the Cartan generators of the gauge group
in four dimensions are given by
\begin{equation}
Q(f)={1\over2}\alpha(f)+F(f),
\label{u1numbers}
\end{equation}
for each complex right--moving fermion $f$.
In the usual notation the 64 worldsheet fermions
in the light--cone gauge are denoted as:
$\psi^\mu, \chi^{1,\dots,6},y^{1,\dots,6}, \omega^{1,\dots,6}$ and
$\overline{y}^{1,\dots,6},\overline{\omega}^{1,\dots,6}$,
$\overline{\psi}^{1,\dots,5}$, $\overline{\eta}^{1,2,3}$,
$\overline{\phi}^{1,\dots,8}$.
Further details
on the formalism and notation used in the free fermionic construction
can be found in the literature \cite{fff, fsu5, slm, alr, lrs, su421,
fknr, fkr, xmap}.

\subsection{Construction of phenomenological models}

Phenomenological free fermionic heterotic--string models were
constructed following two main routes. The first are the so--called
NAHE--based models.
This set of models utilise a set of eight or nine
boundary condition basis vectors. The first five consist of the so--called
NAHE--set \cite{nahe} and are common in all these models. The basis vectors
underlying the NAHE--based models therefore differ by the additional
three or four basis vectors that extend the NAHE--set.

The second route follows from the classification methodology that was
developed in \cite{gkr} for the classification of type II
free fermionic superstrings and adopted in
\cite{fknr, fkr, acfkr, frs} for the
classification of free fermionic
heterotic--string vacua with $SO(10)$ GUT symmetry and its Pati--Salam
\cite{acfkr} and flipped $SU(5)$ \cite{frs} subgroups. The main
difference between the two classes of models is that while the
NAHE--based models allow for asymmetric boundary conditions with respect
to the set of internal fermions $\{ y, \omega\vert {\bar y}, {\bar\omega}\}$,
the classification method only utilises symmetric
boundary conditions. This distinction affects the moduli spaces
of the models \cite{moduli}, which can be entirely fixed in the
former case \cite{cleaver} but not in the later.
On the other hand the classification method enables the systematic
scan of spaces of the order of $10^{12}$ vacua, and led to the
discovery of spinor--vector duality \cite{fkr, svduality} and
exophobic heterotic--string vacua \cite{acfkr}. In this paper,
for reasons that will be clarified below, our discussion is
focused on the NAHE--based models.

\subsubsection{The NAHE set}

The NAHE set \cite{nahe} is a set of five boundary condition basis
vectors $\{ {\bf 1}, S, b_1, b_2, b_3\}$. With `1' indicating Ramond
boundary conditions and `0' indicating Neveu--Schwarz boundary
conditions. The NAHE--set basis vectors are given by:
\beqn
 &&\begin{tabular}{c|c|ccc|c|ccc|c}
 ~ & $\psi^\mu$ & $\chi^{12}$ & $\chi^{34}$ & $\chi^{56}$ &
        $\bar{\psi}^{1,...,5} $ &
        $\bar{\eta}^1 $&
        $\bar{\eta}^2 $&
        $\bar{\eta}^3 $&
        $\bar{\phi}^{1,...,8} $ \\
\hline
\hline
      {\bf 1} &  1 & 1&1&1 & 1,...,1 & 1 & 1 & 1 & 1,...,1 \\
          $S$ &  1 & 1&1&1 & 0,...,0 & 0 & 0 & 0 & 0,...,0 \\
\hline
  ${b}_1$ &  1 & 1&0&0 & 1,...,1 & 1 & 0 & 0 & 0,...,0 \\
  ${b}_2$ &  1 & 0&1&0 & 1,...,1 & 0 & 1 & 0 & 0,...,0 \\
  ${b}_3$ &  1 & 0&0&1 & 1,...,1 & 0 & 0 & 1 & 0,...,0 \\
\end{tabular}
   \nonumber\\
   ~  &&  ~ \nonumber\\
   ~  &&  ~ \nonumber\\
     &&\begin{tabular}{c|cc|cc|cc}
 ~&      $y^{3,...,6}$  &
        $\bar{y}^{3,...,6}$  &
        $y^{1,2},\omega^{5,6}$  &
        $\bar{y}^{1,2},\bar{\omega}^{5,6}$  &
        $\omega^{1,...,4}$  &
        $\bar{\omega}^{1,...,4}$   \\
\hline
\hline
    {\bf 1} & 1,...,1 & 1,...,1 & 1,...,1 & 1,...,1 & 1,...,1 & 1,...,1 \\
    $S$     & 0,...,0 & 0,...,0 & 0,...,0 & 0,...,0 & 0,...,0 & 0,...,0 \\
\hline
${b}_1$ & 1,...,1 & 1,...,1 & 0,...,0 & 0,...,0 & 0,...,0 & 0,...,0 \\
${b}_2$ & 0,...,0 & 0,...,0 & 1,...,1 & 1,...,1 & 0,...,0 & 0,...,0 \\
${b}_3$ & 0,...,0 & 0,...,0 & 0,...,0 & 0,...,0 & 1,...,1 & 1,...,1 \\
\end{tabular}
\label{nahe}
\eeqn
A suitable choice of GGSO phases that preserves
spacetime supersymmetry is given by
\beq
      c\left( \begin{matrix}b_i\cr b_j\cr\end{matrix}\right)~=~
      c\left( \begin{matrix}b_i\cr S\cr\end{matrix}\right) ~=~
      c\left( \begin{matrix}\bone \cr \bone \cr\end{matrix}\right) ~= ~ -1~,
\label{nahephases}
\eeq
where all other GGSO projection coefficients are determined by
modular invariance. The basis vector $S$ is the generator
of spacetime supersymmetry. It merely corresponds to the
Ramond vacuum of the worldsheet fermionic superpartners
of the ten dimensional heterotic--string, and acts as
a spectral flow operator that mixes between the spacetime
fermionic and bosonic states.

The subset
of basis vectors $\{{\bf 1} , S\}$ produces a string vacuum
with $N=4$ spacetime supersymmetry and $SO(44)$ gauge group.
Adding the basis vectors $b_1$ and $b_2$ reduces the
$N=4$ spacetime supersymmetry to $N=1$, where each of these
vectors on its own reduces $N=4$ to $N=2$, and their
combined action reduces $N=4$ to $N=1$. The characteristic
of these two vectors is that their overlap with the basis
vector $S$ yields $S\cdot b_{1,2}= 2$. Thus, any additional
vector that satisfies this overlap with the basis vector
$S$ reduces the number of supersymmetries from $N=4$ to
$N=2$. An example at hand is the basis vector $b_3$ of the
NAHE--set. However, the additional breaking induced by
any additional basis vector with this property either
preserves $N=1$ supersymmetry or reduces it further
to $N=0$.

 This type of breaking therefore produces
a type of non supersymmetric string vacua that
follow the chain
$N=4\rightarrow N=2\rightarrow N=1\rightarrow N=0$.
One characteristic of this type of spacetime supersymmetry
breaking is that the breaking will a priori not be
family universal. The reason being that the chiral families
arise in the free fermionic models from the three sectors $b_1$,
$b_2$ and $b_3$, and this type of breaking necessarily
violates the cyclic permutation symmetry among the
three sectors $b_1$, $b_2$ and $b_3$.

An example,
where this cyclic permutation symmetry is instrumental
in producing a family universal structure, is the
supersymmetry breaking with a family universal
anomalous $U(1)$ of \cite{u1abreaking}.
The basis vectors $b_1$, $b_2$ and $b_3$ reduce the
$SO(44)$ gauge symmetry to $SO(10)\times SO(6)^3\times E_8$.
The gauge bosons that produce this gauge symmetry are obtained
from the Neveu--Schwarz (NS) sector and the sector
$Z={\bf 1} + b_1 + b_2 + b_3$, where the NS produces the
vector bosons of $SO(10)\times SO(6)^3\times SO(16)$ and the
sector $Z$ complements the $SO(16)$ group factor to $E_8$.

The NAHE set basis vectors $b_1$, $b_2$ and $b_3$
correspond to the three twisted sectors of the $Z_2\times Z_2$
toroidal orbifold. Each of these twisted sectors produces
sixteen multiplets in the spinorial ${\bf 16}$ representation
of $SO(10)$ to give a total of forty eight chiral generations.
The correspondence of the quasi--realistic
free fermionic models with $Z_2\times Z_2$ orbifold has been
amply discussed in the literature \cite{z2z2}. While the
dictionary between specific models in the two approaches
may be elusive, it is anticipated that for every
model in one formalism there exist a representation in the
alternative formalism and this should hold, at least
for the $Z_2\times Z_2$ orbifolds, and higher order orbifolds may
have a fermionic representation as well \cite{kakutye}.

\subsubsection{Beyond the NAHE--set}
The construction of the semi--realistic free fermionic models
proceeds by adding three or four additional basis vectors
to the NAHE--set.
The function of the additional basis
vectors is to reduce the forty eight spinorial ${\bf 16}$ multiplets
to three chiral generations, and at the same time to
reduce the $SO(10)$ GUT symmetry to one
of its subgroups:
\begin{enumerate}[label=\roman*]
\item $SU(5)\times U(1)$ (FSU5) \cite{fsu5};
\item $SU(3)\times SU(2)\times U(1)^2$ (SLM) \cite{slm};
\item $SO(6)\times SO(4)$ (PS) \cite{alr};
\item $SU(3)\times U(1)\times SU(2)^2$ (LRS) \cite{lrs};
\item $SU(4)\times SU(2)\times U(1)$ (SU421) \cite{su421}.
\end{enumerate}
The first four cases produced viable three generation models,
whereas in the last case it was shown that phenomenologically viable
models cannot be constructed \cite{su421, su421class}.
The additional basis vectors may each preserve or break
the $SO(10)$ symmetry. Basis vectors that preserve the $SO(10)$ symmetry are
typically denoted by $b_i$ with $(i=4, 5, \dots)$, whereas
those that break the $SO(10)$ symmetry are denoted by
$\{\alpha, \beta, \gamma\}$.
The overlap of the additional basis vectors with the supersymmetry generator
basis vector $S$ determine the type of possible supersymmetry breaking.
Thus, in the cases with $S\cdot b_i= 2 $ the pattern of supersymmetry
breaking is similar to the spacetime supersymmetry breaking
discussed above, namely it follows the chain
$N=4\rightarrow N=2\rightarrow N=1\rightarrow N=0$.

An alternative is to use a basis vector with
$S\cdot a_i=0$, where $a_i$ may, or may not, break the $SO(10)$
GUT symmetry. This type of supersymmetry breaking differs, however,
from the one discussed above in that it induces the breaking pattern
$N=4\rightarrow N=0$. A general rule to construct vacua that preserve
$N=1$ spacetime supersymmetry is to impose $\cc{S}{v_i}=-\delta_{v_i}$,
where $v_i$ is any basis vector \cite{toward}. Thus, relaxing this
constraint would generically result in broken spacetime supersymmetry.
Breaking spacetime supersymmetry with the additional basis
vectors $a_i$ would generically also not affect the spectrum
arising from the sectors $b_i$ that produce the chiral generations,
but may affect their superpartners as those are obtained
from the sectors $S+b_i$. As we will show with an explicit
example model in section \ref{explicitmodel},
in this case the effect of the projection is to select
different components of the underlying $N=4$ multiplets.

\subsubsection{The classification set}\label{classiset}
In this paper our focus will be on non--supersymmetric NAHE--based
free fermionic models. For completeness we discuss the
construction of such models by using the classification
methods of \cite{fknr, acfkr, frs, fkr}.
In this approach the set of basis vectors is fixed and
a large number of string models, of the order of $10^{12}$
vacua, is spanned by enumerating the independent GGSO
projection coefficients. In this manner large spaces
of string models with $SO(10)$ \cite{fknr}, $SO(6)\times SO(4)$
\cite{acfkr}, $SU(5)\times U(1)$ \cite{frs}, and
$SU(4)\times SU(2)\times U(1)$ \cite{su421class},
have been explored. A subset of basis vectors
that respect the $SO(10)$ symmetry is given by
the set of 12 basis vectors
$
V=\{v_1,v_2,\dots,v_{12}\},
$
where
\begin{eqnarray}
v_1=1&=&\{\psi^\mu,\
\chi^{1,\dots,6},y^{1,\dots,6}, \omega^{1,\dots,6}| \nn\\
& & ~~~\bar{y}^{1,\dots,6},\bar{\omega}^{1,\dots,6},
\bar{\eta}^{1,2,3},
\bar{\psi}^{1,\dots,5},\bar{\phi}^{1,\dots,8}\},\nn\\
v_2=S&=&\{\psi^\mu,\chi^{1,\dots,6}\},\nn\\
v_{2+i}=e_i&=&\{y^{i},\omega^{i}|\bar{y}^i,\bar{\omega}^i\}, \
i=1,\dots,6,\nn\\
v_{9}=b_1&=&\{\chi^{34},\chi^{56},y^{34},y^{56}|\bar{y}^{34},
\bar{y}^{56},\bar{\eta}^1,\bar{\psi}^{1,\dots,5}\},\label{basis}\\
v_{10}=b_2&=&\{\chi^{12},\chi^{56},y^{12},y^{56}|\bar{y}^{12},
\bar{y}^{56},\bar{\eta}^2,\bar{\psi}^{1,\dots,5}\},\nn\\
v_{11}=z_1&=&\{\bar{\phi}^{1,\dots,4}\},\nn\\
v_{12}=z_2&=&\{\bar{\phi}^{5,\dots,8}\}.\nn
\end{eqnarray}
Additional vectors are added to the set given in (\ref{basis})
to construct vacua with $SO(10)$ subgroups \cite{acfkr,frs}.
In the notation of eq. (\ref{basis})
the worldsheet fermions appearing in the
curly brackets have periodic boundary conditions, whereas
all other worldsheet fermions are anti--periodic.
The entries in the matrix of GGSO phases $\cc{v_i}{v_j}$
with $i>j$ then span the space of string vacua. Additional
constraints that are imposed on the string vacua, like the existence
of spacetime supersymmetry leave 40 independent phases of the
original 66. One can then resort to a complete \cite{fkr} or
statistical sampling\footnote{We note that analysis of large sets
of string vacua has also been carried out by other groups
\cite{statistical}.} of the total space
\cite{fknr},
and classify the models by their twisted matter spectrum.
The classification is facilitated by expressing the GGSO
projections in algebraic form \cite{fknr,fkr}.
The analysis of the entire spectrum of the string models
is computerised and vacua with specific phenomenological
characteristics can be fished our from the larger space of models.

In terms of spacetime supersymmetry breaking, as with the NAHE--set
based models the spacetime supersymmetry generator is the
basis vector $S$. The subset $\{{\bf 1}, S\}$ gives rise
to $N=4$ spacetime supersymmetry, which is broken by
$b_1$ and $b_2$ to $N=2$ spacetime supersymmetry and their
combined action breaks $N=4\rightarrow N=1$.
As with the NAHE--based models imposing $\cc{S}{v_i}=-\delta_{v_i}$
ensures the preservation of $N=1$ supersymmetry. Projecting
the remaining supersymmetry in this model
is obtained by relaxing this condition. Furthermore, the
basis vectors $\{e_i, z_1, z_2\}$ satisfy $S\cdot e_i= 0$ and
$S\cdot z_i= 0$. These basis vectors therefore act as projectors
on the $S$--sector. These basis vectors can be used to project all
the states from the $S$--sector and hence induce the breaking
$N=4\rightarrow N=0$ spacetime supersymmetry.

\section{Tachyons in the free fermionic semi-realistic models }\label{tachfree}

String models, in general, and heterotic--string models in particular,
generically give rise to tachyonic states in their spectra.
This can be seen from eq. (\ref{virasorocond}).
Any sector that satisfies
\beq
M_L^2~<~ -{1\over 2} ~~~{\rm  and} ~~~M_R^2~<~-1
\label{tachyonicsectors}
\eeq
may produce tachyonic physical states. Tachyonic states can be obtained
by acting on the vacuum with fermionic oscillators.
They satisfy the level matching condition and survive all the
GGSO projections. Their presence in the physical spectrum indicates
the instability of the string vacuum.
The existence of spacetime supersymmetry guarantees that all
tachyonic states are projected out. The situation is altered
if supersymmetry is broken to $N=0$ spacetime supersymmetry
by projecting all the states from the $S$--sector.
One then has to check in each model whether tachyonic
states exist.

The existence of non--supersymmetric
non--tachyonic string vacua has been known since the mid--eighties
\cite{aggmv}. The gauge symmetry of this model
is $SO(16)\times SO(16)$, and its non--perturbative extension
was considered in \cite{ft1}. In the free fermionic formalism
the model is constructed by the set of boundary condition basis
vectors $\{{\bf 1}, S, X, I\}$ where
$X= \{{\bar\psi}^{1,\cdots,5}, {\bar\eta}^{1,2,3}$\} and
$I=\{{\bar\phi}^{1,\cdots,8}\}$. In ten dimensions
the choice of the GGSO phase
$\cc{X}{I}=\pm1$  yields either the supersymmetric
$E_8\times E_8$, or the non--supersymmetric $SO(16)\times SO(16)$,
heterotic--string. This is necessarily the case in ten dimensions
because the supersymmetry generator is given by $S={\bf1}+X+I$
and  therefore the projections on the three sectors are correlated.
In the four dimensional models the same phase can be used to reduce
the gauge symmetry from $E_8\times E_8$ to $SO(16)\times SO(16)$
without breaking supersymmetry. The same vacua can be
constructed in the orbifold representation and can be connected
by interpolations \cite{ft2}. Hence, the supersymmetric and
non--supersymmetric vacua exist on the boundary of the same moduli
space.

Our interest in this paper is in the tachyonic states arising in the
semi--realistic models. It is instructive to examine the case of the
non--supersymmetric $SO(16)\times SO(16)$ model first. In the four
dimensional model supersymmetry may be broken from $N=4\rightarrow N=0$
by the $I$ or $X$ projections. The only sector that may produce tachyons
in this model is the NS sector. The tachyonic states arising in this
model are obtained by acting on the non--degenerate vacuum with a
right--moving oscillator, and satisfy the level matching condition
with $M_L^2=M_R^2= -1/2$. These tachyonic states are, however, projected
out by the $S$ projection, which is given by
\beq
{\rm e}^{i\pi S\cdot F_{NS}}\vert t\rangle_{NS} = \delta_S\vert t\rangle_{NS}.
\label{snsprojection}
\eeq
As there are no oscillators acting on the left--moving vacuum
in the tachyonic untwisted state, and the basis vector $S$ is
blind to the right--moving oscillators, the left--hand side
of eq. (\ref{snsprojection}) is positive. On the other hand
$\delta_S=-1$ because the spacetime fermions $\psi^\mu$ are
periodic in $S$.
The mismatch
between the two sides of eq. (\ref{snsprojection})
entails that the untwisted NS tachyons are projected out.
This argument extends to any free fermionic model that
contains the basis vector $S$. We conclude that
in any non--supersymmetric free fermionic
that includes the $S$--sector the untwisted
tachyons are always projected out,
irrespective of the choice of the SUSY projecting phases.

The four dimensional $SO(16)\times SO(16)$ non--supersymmetric
heterotic--string is therefore tachyon free. However,
in this string vacuum the only sector that may give rise
to tachyonic states is the untwisted sector. This is not
the case in the semi--realistic free fermionic models. The
three generation free fermionic models generically give rise
to an abundance of sectors that may a priori give rise
to tachyonic states. The reason is that the additional
basis vectors that are required to reduce the number of families,
break down the internal degrees of freedom into
small units. This is exemplified by the set of
vectors in eq. (\ref{basis}), which is used as the
basis for the classification of symmetric fermionic
$Z_2\times Z_2$ orbifolds.
The basis vectors $e_i$ as well as their $e_i+e_j ~i\ne j$ and
$e_i+e_j+e_k ~i\ne j\ne k$ combinations
may produce physical tachyonic states
by acting on the non degenerate vacuum
with a right--moving NS fermionic oscillator;
similarly the sectors $z_\ell$, $z_\ell+e_i$, $z_\ell+e_i+e_j,~i\ne j$
$z_\ell+e_i+e_j+e_k,~i\ne j\ne k$ may produce tachyonic states.
In total there are therefore 123 sectors, in addition to the
NS--sector, that may produce tachyons in these models.

This renders futile a systematic classification of non--supersymmetric
non--tachyonic semi--realistic vacua along the lines of \cite{acfkr, frs}. The
situation in NAHE--based models is similar.
Typical models contain an abundance of sectors that may a priori
produce tachyons. Furthermore, models that utilise fractional boundary
conditions may contain additional tachyon producing sectors
in which a fermionic oscillator with rational boundary
conditions may act on the non--degenerate vacuum.
A detailed sector by sector analysis is therefore required.
A systematic procedure to extract the tachyon free vacua is
provided by performing a $q$--expansion of the partition function
\cite{abeldienes}. However, this method loses the detailed
information on the structure of the string spectrum.
The construction of non--supersymmetric vacua with quasi--realistic
features therefore requires a detailed model by model analysis.
One may then envision the existence of models in which the
number of tachyonic producing sectors is restricted.

The best case scenario would be a model in which
the only tachyon producing sector is the NS--sector.
In this case we are guaranteed that tachyons do not exist
in the physical spectrum. However, a model with this property
has not been found to date. The next best case scenario is
a model that gives rise only to one type of tachyon producing
sectors. Existence of a model with this characteristic
may depend on further detailed phenomenological properties
of the string vacua. For example, we were not able to
find such a model in the class of NAHE--based free fermionic
models with reduced Higgs spectrum \cite{cleaver}, whereas the
class of left--right symmetric models \cite{lrs} did
produce a model with the desired property.
The set of boundary condition basis vectors, beyond the NAHE--set,
generating the string vacuum is given by
\beqn
 &\begin{tabular}{c|c|ccc|c|ccc|c}
 ~ & $\psi^\mu$ & $\chi^{12}$ & $\chi^{34}$ & $\chi^{56}$ &
        $\bar{\psi}^{1,...,5} $ &
        $\bar{\eta}^1 $&
        $\bar{\eta}^2 $&
        $\bar{\eta}^3 $&
        $\bar{\phi}^{1,...,8} $ \\
\hline
\hline
 ${\alpha}$  &  0 & 0&0&0 & 1~1~1~0~0 & 0 & 0 & 0 &1~1~1~1~0~0~0~0 \\
 ${\beta}$   &  0 & 0&0&0 & 1~1~1~0~0 & 0 & 0 & 0 &1~1~1~1~0~0~0~0 \\
 ${\gamma}$  &  0 & 0&0&0 &
${1\over2}$~${1\over2}$~${1\over2}$~0~0&${1\over2}$&${1\over2}$&${1\over2}$
&0~${1\over2}$~${1\over2}$~${1\over2}$~${1\over2}$~${1\over2}$~${1\over2}$~0 \\
\end{tabular}
   \nonumber\\
   ~  &  ~ \nonumber\\
   ~  &  ~ \nonumber\\
     &\begin{tabular}{c|c|c|c}
 ~&   $y^3{y}^6$
      $y^4{\bar y}^4$
      $y^5{\bar y}^5$
      ${\bar y}^3{\bar y}^6$
  &   $y^1{\omega}^5$
      $y^2{\bar y}^2$
      $\omega^6{\bar\omega}^6$
      ${\bar y}^1{\bar\omega}^5$
  &   $\omega^2{\omega}^4$
      $\omega^1{\bar\omega}^1$
      $\omega^3{\bar\omega}^3$
      ${\bar\omega}^2{\bar\omega}^4$ \\
\hline
\hline
$\alpha$& 1 ~~~ 1 ~~~ 1 ~~~ 0  & 1 ~~~ 1 ~~~ 1 ~~~ 0  & 1 ~~~ 1 ~~~ 1 ~~~ 0 \\
$\beta$ & 0 ~~~ 1 ~~~ 0 ~~~ 1  & 0 ~~~ 1 ~~~ 0 ~~~ 1  & 1 ~~~ 0 ~~~ 0 ~~~ 0 \\
$\gamma$& 0 ~~~ 0 ~~~ 1 ~~~ 1  & 1 ~~~ 0 ~~~ 0 ~~~ 0  & 0 ~~~ 1 ~~~ 0 ~~~ 1 \\
\end{tabular}
\label{model3}
\eeqn
This model gives rise only to one type of tachyon producing sectors with
\beq
\alpha_L^2 ~ = ~ 2 ~~~\&         ~~~ \alpha_R^2 ~=~ 6 ~~~\Rightarrow~~~
                                 N_R ~=~ 0\label{ar6nr0}
\eeq
The supersymmetric version of this model was presented in \cite{lrs}
with the set of GGSO phases given by
\begin{equation}
{\bordermatrix{
          &{\bf 1}& S & &{b_1}&{b_2}&{b_3}& &{\alpha}&{\beta}&{\gamma}\cr
       {\bf 1}&~~1&~~1 & & -1   &  -1 & -1  & & ~~1     & ~~1   & ~~i   \cr
             S&~~1&~~1 & &~~1   & ~~1 &~~1  & &  -1     &  -1   &  -1   \cr
	      &   &    & &      &     &     & &         &       &       \cr
       {  b_1}& -1& -1 & & -1   &  -1 & -1  & &  -1     &  -1   & ~~i   \cr
       {  b_2}& -1& -1 & & -1   &  -1 & -1  & &  -1     &  -1   & ~~i   \cr
       {  b_3}& -1& -1 & & -1   &  -1 & -1  & &  -1     & ~~1   & ~~i   \cr
	      &   &    & &      &     &     & &         &       &       \cr
      {\alpha}&~~1& -1 & &~~1   & ~~1 &~~1  & & ~~1     & ~~1   & ~~1   \cr
       {\beta}&~~1& -1 & & -1   &  -1 &~~1  & &  -1     &  -1   &  -1   \cr
      {\gamma}&~~1& -1 & &~~1   &  -1 &~~1  & &  -1     &  -1   & ~~1   \cr}}.
\label{phasesmodel3}
\end{equation}
The full mass spectrum of this model together with the cubic level
superpotential
was presented in \cite{lrs}. The modification
\beq
\cc{S}{\alpha}=-1\rightarrow+1
{}~~~{\rm and}~~~
\cc{S}{\beta}=-1\rightarrow+1
\label{modifiedsxphases}
\eeq
projects the remaining gravitino and induces $N=1\rightarrow N=0$.
It can be checked that all the tachyonic states are projected out in this
model. Furthermore, it can be verified that making the modification
\beq
\cc{S}{\alpha}=-1\rightarrow+1
{}~~~{\rm and}~~~
\cc{S}{\beta}=-1\rightarrow-1
\label{modifiedsxtphases}
\eeq
{\it i.e.}~modifying only $\cc{S}{\alpha}$ but not $\cc{S}{\beta}$
results in a model that contains tachyonic states. The reason is that
in this model all the sectors that may produce tachyons appear with the
combination $m(\alpha+\beta)$, where $m=0,1$. Hence, with the modification
given by eq. (\ref{modifiedsxphases}) the $S$--projection on the tachyonic
sectors is the same as in the corresponding supersymmetry
preserving choice given in eq. (\ref{phasesmodel3}), whereas with the
modification given by (\ref{modifiedsxtphases}) the $S$--projection  in some
sectors is
modified in comparison to the supersymmetric model and
some tachyonic states are not projected out. We note that the
construction of tachyonic free semi--realistic vacua is highly
nontrivial. In the next section we discuss the tachyon free
model in some detail.

\section{An explicit tachyon-free model}\label{explicitmodel}
We consider the model defined by the set of basis vectors
\begin{eqnarray*}
1 &=&\{\psi^\mu,\
\chi^{1,\dots,6},y^{1,\dots,6}, \omega^{1,\dots,6}|
\bar{y}^{1,\dots,6},\bar{\omega}^{1,\dots,6},
\bar{\eta}^{1,2,3},
\bar{\psi}^{1,\dots,5},\bar{\phi}^{1,\dots,8}\},\nn\\
S &=& \{\psi^{\mu},\chi^{1,..,6}\}\\
b_{1} &=& \{\psi^{\mu},\chi^{1,2},
%
y^{3,...,6}|\overline{y}^{3,...,6},\overline{\psi}^{1,...,5},\overline{\eta}^{1}\}\\
b_{2} &=& \{\psi^{\mu},  \chi^{3,4},
%
y^{1,2},\omega^{5,6}|\overline{y}^{1,2},\overline{\omega}^{5,6},\overline{\psi}^{1,...,5},\overline{\eta}^{2}  \}\\
b_{3} &=& \{\psi^{\mu}, \chi^{5,6}, \omega^{1,...,4}|
\overline{\omega}^{1,...,4}, \overline{\psi}^{1,...,5}, \overline{\eta}^{3}
\}\\
b_{4} &=& \alpha \\
%
&=&\{y^{1,...,6},\omega^{1,...,6}|\overline{\omega}^{1},\overline{y}^{2},\overline{\omega}^{3},\overline{y}^{4,5},\overline{\omega}^{6},\overline{\psi}^{1,2,3},\overline{\phi}^{1,...,4}\}\\
b_{5} &=& \beta \\
&=&\{y^{2},\omega^{2}, y^{4}, \omega^{4}
|\overline{y}^{1,...,4},\overline{\omega}^{5}, \overline{y}^{6},
\overline{\psi}^{1,2,3},\overline{\phi}^{1,...,4}\}\\
b_{6} &=& \gamma\\
&=&\{y^{1}, \omega^{1}, y^{5}, \omega^{5} | \overline{\omega}^{1,2},
\overline{y}^{3},\overline{\omega}^{4},\overline{y}^{5,6},
\overline{\psi}^{1,2,3}=\frac{1}{2}, \overline{\eta}^{1,2,3}=\frac{1}{2},
\overline{\phi}^{2,...,7}=\frac{1}{2}\}\\
\end{eqnarray*}
with the set of GGSO phases given by
\begin{center}
$
\bordermatrix{~ & 1 & S &b_{1}&b_{2}&b_{3}&\alpha&\beta&\gamma \cr
              1 & \,\,\,\,1&1&-1&-1&-1&1&\,\,\,\,1&\,\,\,\,i \cr
              S
&\,\,\,\,1&\,\,\,\,1&\,\,\,\,1&\,\,\,\,1&\,\,\,\,1&\,\,\,\,1&\,\,\,\,1&-1  \cr
              b_{1} &-1&-1&-1&-1&-1&-1&-1&\,\,\,\,i\cr
              b_{2} &-1&-1&-1&-1&-1&-1&-1&\,\,\,\,i\cr
              b_{3} &-1&-1&-1&-1&-1&-1&\,\,\,\,1&\,\,\,\,i\cr
              \alpha
%
&\,\,\,\,1&\,\,\,\,1&\,\,\,\,1&\,\,\,\,1&\,\,\,\,1&\,\,\,\,1&\,\,\,\,1&\,\,\,\,1\cr
              \beta &\,\,\,\,1&\,\,\,\,1&-1&-1&-1&-1&-1& -1\cr
              \gamma &\,\,\,\,1&-1&\,\,\,\,1&-1&\,\,\,\,1&-1&-1&\,\,\,\,1 \cr}.
$
\end{center}
 This is a 3 generation model, with one generation appearing in each of the
twisted sectors $b_1, b_2$ and $b_3$. The full spectrum can be found in the
table of appendix  \ref{sec:Sectors}, with the exception of the gauge bosons
which have been omitted in the interest of space. It is sufficient to state
that the gauge group is
$$\underbrace{SU(3)_C \times U(1)_C \times SU(2)_L \times SU(2)_R
\times\prod_{i=1}^6 U_i}_{\text{observable sector}} \times
\underbrace{SU(3)_{H_1} \times SU(3)_{H_2}\times\prod_{j=7}^{10}
U_j}_{\text{hidden sector}}.$$
The notation for the table is the following:  The first column describes if the
states correspond to spacetime bosons or spacetime fermions and specifically
for $b_i$ the type of particle. The second column is the name of the sector.
The third column gives the dimensionality of the states under $SU(3)_C \times
SU(2)_L \times SU(2)_R$ and the fourth the charges of the observable $U(1)$s.
Columns 5 and 6 describe the hidden sector. The only charges appearing in the
table that do not have a self--evident name are:
\begin{eqnarray}
Q_C&=&
Q_{\overline{\psi}^{1}}+Q_{\overline{\psi}^{2}}+Q_{\overline{\psi}^{3}}~,\nonumber\\
Q_8&=&
Q_{\overline{\phi}^{2}}+Q_{\overline{\phi}^{3}}+Q_{\overline{\phi}^{4}}~,\nonumber\\
Q_9&=&
Q_{\overline{\phi}^{5}}+Q_{\overline{\phi}^{6}}+Q_{\overline{\phi}^{7}}~.
\end{eqnarray}
To avoid writing fractional numbers all the charges in the table have been
multiplied by $4$. Finally, for every state the CPT conjugate is also
understood to be in the spectrum and has not been written explicitly.
Lastly, we comment that the states contain discrete charges 
corresponding to the action of real fermions that are 
not shown in the table. For example, the first three states 
from the Neveu--Schwarz sector are obtained by acting on the 
vacuum with two right--moving real fermions and are neutral
under the gauge symmetry of the model.
The weak hypecharge in the model is given by 
$$U(1)_Y=\frac{1}{3}U(1)_C+T_{3_R},$$
where $T_{3_R}$ is the diagonal generator of $SU(2)_R$.
The symmetry breaking to the Standard Model 
gauge group may be induced by a VEV for one of the Standard Model
singlet scalar fields
in the $(1,1,2)$ representation from the sectors $S+b_j$. 
This model exhibits many interesting features regarding supersymmetry. Firstly,
we observe that the model is manifestly non-supersymmetric. The gravitino and
the gaugini are projected out and there is a clear mismatch between the number
of states in the $0$ and $S$ sectors. Furthermore, there are eight sectors with
only scalars and the sectors that contain the would-be superpartners 
are massive. These are

\begin{center}
\begin{equation}
\begin{tabular}{cc}
$\beta+\gamma$,&   $\beta+3\gamma$,\\ $\alpha+\gamma$,&$\alpha+3\gamma$,\\
$1+b_1+b_2+b_3+\beta+\gamma$,&$1+b_1+b_2+b_3+\beta+3\gamma$,\\
$1+b_1+b_2+b_3+\alpha+\gamma$,&$1+b_1+b_2+b_3+\alpha+3\gamma$.
\end{tabular}
\label{badsectors}
\end{equation}
\end{center}

Such sectors would not remain in the spectrum in the supersymmetric choice of
phases. The reason is that the spacetime supersymmetry generator in the
supersymmetric model is the basis vector $S$, {\it i.e.} for 
a given sector $\rho\in \Xi$, the supersymmetric superpartners are
obtained from the sector $S+\rho$. All the sectors in eq. (\ref{badsectors})
have $(\rho)_L^2=4$, whereas $(S+\rho)_L^2=8$, {\it i.e.} in these
sectors the would-be superpartners are massive. In the 
supersymmetric vacua the states from the sectors in eq. (\ref{badsectors})
are necessarily projected out, as they break supersymmetry explicitly.
However, once supersymmetry is broken they may appear 
in the spectrum, as is seen in our model. 
It is a highly non-trivial task to find a model with 3 generations in
which sectors of these type, that only appear when supersymmetry is broken,
contain no tachyons, but this model provides exactly such an example. We
collectively refer to all the sectors mentioned in this paragraph as sectors in
which supersymmetry is ``badly broken".

On the other hand, there are (pairs of) sectors that are completely
supersymmetric. This is due to the modification (\ref{modifiedsxphases}) not
affecting the GGSO projections in any sectors where none of the vectors $S,
\alpha$ or $\beta$ appear. Therefore such sectors will be identical to the
corresponding sectors of the supersymmetric model. Nonetheless, for some of
these sectors to remain supersymmetric as claimed above, the superpartners
should be unchanged as well, or at least the effect must be (at most) a change
in the $R$--charges of the superpartners. Sectors $b_i$ and $1+b_i+b_j+2\gamma$
are of this type.

Finally, there are sectors that do not fit any of the above categories. In
these sectors the number of bosons and fermions is the same, but on the other
hand some of the gauge charges of these states are different which in principle
prevents us from grouping them together into supermultiplets. 
Most of the sectors are of this type. We use the term sectors in which
supersymmetry is ``nicely broken" when referring to this case.

Thus, while supersymmetry is broken, some segments of the
string vacuum still respect the underlying supersymmetric degeneracy.
This is in accordance with the findings in \cite{fkp}, which showed
that the partition function of string vacua with spontaneously
broken supersymmetry can be divided into several orbits,
some of which preserve the original supersymmetry.

Furthermore,
we would like to comment
in our model the fermionic states from the sectors
$b_1$, $b_2$ and $b_3$, as well as the bosonic
states from the NS--sector, are not affected by the GGSO phases
that project the gravitino and gaugini from the $S$--sector,
and therefore break spacetime supersymmetry.
Therefore, the untwisted scalar
states of our non--supersymmetric model as well as the fermionic
states from the
sectors $b_1$, $b_2$ and $b_3$ are identical to those in the
corresponding supersymmetric model. Consequently,
the leading twisted--twisted--untwisted couplings in
the non--supersymmetric model, which are obtained by using the
methods developed in \cite{kln}, are identical to those of
the supersymmetric model.
The model generated by eqs. (\ref{model3},\ref{phasesmodel3})
contains electroweak doublet scalar representations
from the twisted sectors that may be used as Higgs doublets.
However, in this model the untwisted Higgs bi--doublets,
which couple at leading order to the twisted sector states,
are projected out and consequently the leading
mass term which is identified with the top mass is
absent. Other LRS \cite{lrs}  models, as well as
the FSU5 \cite{fsu5}, PS \cite{alr} and SLM \cite{fny,slm} models,
do contain the untwisted Higgs doublets and in those cases a
leading top mass term is obtained.

It is also worth noting that even for non-supersymmetric models the
cosmological constant can be exponentially suppressed. As discussed in
\cite{abeldienes}, this can be achieved if the massless spectrum has an equal
number of bosons and fermions (irrespectively of their charges). Even though
our model is not of this type and will therefore have an unsuppressed vacuum
energy, our construction hints at how one might go about achieving such a goal.
It is clear for example, that we do not have to worry about sectors that either
respect supersymmetry or in which supersymmetry is nicely broken.

On the other hand, sectors that badly break supersymmetry will have to be
carefully engineered. There are a few ways one might go about such a task. For
example, one might entertain the idea that the addition of further basis
vectors could project such sectors out of the spectrum. The biggest problem
with this approach is that the removal of the gaugini from the S  sector, even
if some fermions transforming in a different than the adjoint representation
are preserved, will create a mismatch of states in the S and NS sectors turning
them into sectors that break supersymmetry badly; and it is impossible to
project out the NS sector no matter what basis vectors are added. It is
\textit{a priori} possible that further basis vectors will remove exactly the
correct number of bosons from the NS sector to match the remaining fermionic
states in the S sector, but this method seems unnecessarily restricting.

An approach providing more freedom is to aim for an equality in the number of
bosons and fermions not in each sector, but among different sectors. To cancel
the surplus of bosons from the NS sector this would imply the existence of
surviving fermionic states in different sectors, the bosonic counterpart of
which has been projected out.
The model presented in this section has an overall mismatch of 
bosonic and fermionic degrees of freedom and therefore does 
not satisfy this condition.  
Finding a semi-realistic model of suppressed cosmological constant
appears to be very challenging, but it is of great interest as well and we hope
to report on such constructions in a future publication.

\section{Split SUSY models }\label{split}
In this section we briefly discuss string models with a split supersymmetry
structure. The basic idea is to use two basis vectors to generate space--time
supersymmetry. We recall from section \ref{review} that in the semi--realistic
free fermionic models the supersymmetry generators arise from
the basis vector $S$.
The aim in split supersymmetry string models is to construct two basis vectors
that produce supersymmetry generators.
A particular aim is then to construct models
in which gaugini are obtained from one generator, whereas those
of the second generator are projected out, as well as
the scalar superpartners of the twisted matter fermionic states.
Our construction proceeds by keeping our previous
basis vector $S=\{\psi^{1,2}, \chi^{1, \dots, 6}\}\equiv S_1$. A second
supersymmetry generator is given by
\beq
S_2=\{ \psi^{1,2}, \chi^{1,2}, \omega^{3,4}, \omega^{5,6}\}.
\label{s2}
\eeq
The basis vectors $b_1$ and $b_2$ of the NAHE--set eq. (\ref{nahe})
are added as well as the basis vectors ${\bf 1}$ and $X$, which is
used to project the supersymmetric generators from $S_1$,
as discussed in section \ref{tachfree}. Shift basis vectors similar
to the $e_i$ basis vectors of eq. \ref{basis} can be added,
and variations that include the basis vector $I$ of section
\ref{review}. We consider the set of six basis vectors
given by
\begin{eqnarray}
v_1=1&=&\{\psi^\mu,\
\chi^{1,\dots,6},y^{1,\dots,6}, \omega^{1,\dots,6}| \nn\\
& & ~~~\bar{y}^{1,\dots,6},\bar{\omega}^{1,\dots,6},
\bar{\eta}^{1,2,3},
\bar{\psi}^{1,\dots,5},\bar{\phi}^{1,\dots,8}\},\nn\\
v_2=S_1&=&\{\psi^\mu,\chi^{1,\dots,6}\},\nn\\
v_3=S_2&=&\{\psi^\mu,\chi^{1,2},\omega^{3,\dots, 6}\},\label{basiss1s2}\\
v_4=b_1&=&\{\chi^{34},\chi^{56},y^{34},y^{56}|\bar{y}^{34},
\bar{y}^{56},\bar{\eta}^1,\bar{\psi}^{1,\dots,5}\},\nn\\
v_5=X&=&\{\bar{\eta}^{1,2,3},\bar{\psi}^{1,\dots,5}\},\nn\\
v_6=I&=&\{\bar{\phi}^{1,\dots,8}\}.\nn
\end{eqnarray}
with the set of GGSO phases given by
\begin{equation}
{\bordermatrix{
            &{\bf 1}& S_1 & S_2   &b_1  &  X  &   I  \cr
     {\bf 1}& -1    & -1  & -1   &  -1 & -1  &  -1   \cr
         S_1& -1    & -1  & -1   & ~~1 &~~1  & ~~1   \cr
         S_2& -1    &~~1  & -1  & ~~1 & -1  &   -1   \cr
     {  b_1}& -1    & -1  & -1   &  -1 &~~1  &  -1   \cr
     {  X  }& -1    &~~1  & -1   &  -1 &~~1  &  -1   \cr
         {I}& -1    &~~1  & -1  &  -1 &  -1  & ~~1   \cr}}.
\label{phasesmodels1s2}
\end{equation}
In this model the NS--sector is the only sector that
produces spacetime vector bosons. Hence the
gauge symmetry in four dimensions is
$SO(8)\times SO(4) \times SO(4)\times SO(12)\times SO(16)$.
The sector $b_1$ gives rise to spacetime fermions in the spinor
and anti--spinor representations of $SO(12)$. The supersymmetry
generators of $S_1$ are projected out, whereas the gaugini
from $S_2$ are retained. The model retains the
scalar superpartners from the sector $S_2+b_1$,
and projects those from the sector $S_1+b_1$.
Our general aim in the construction of models with
split supersymmetry is to construct models that retain
the gaugini and spacetime fermions from $S_2$ and $b_1$,
while projecting the gaugini (and hence the gravitini) from
$S_1$, as well as the superpartners from the sectors
$S_1+b_1$ and $S_2+b_1$. However, variations of the model
in eqs. (\ref{basiss1s2},\ref{phasesmodels1s2}), including adding
the $e_i$ projectors of eq. (\ref{basis}) did not yield the desired
result. The models in which supersymmetry is entirely broken,
{\it i.e.} those in which the supersymmetry generators from
$S_1$ as well as $S_2$ are projected out, typically contain
tachyons. We then face similar situation to the one
discussed in section \ref{classiset}.

\section{Conclusions}\label{conclusions}

The observation of a scalar resonance compatible with the
electroweak Higgs doublet reinforces the hypothesis that the
Standard Model provides a viable parameterisation of
all sub--atomic data up to the Planck scale. Synthesis
of gravity with the sub--atomic interactions necessitates
a departure from the local point particle idealisation
of quantum field theories, which underly the Standard Model.
The most developed framework to explore the gauge--gravity augmentation is
offered by string theory. Detailed
phenomenological models that incorporate the salient
feature of the Standard Model have been constructed.
These detailed phenomenological constructions
contain a new symmetry, {\it i.e.} $N=1$ spacetime
supersymmetry, which has not been observed to date
in experiments.

A vital question is therefore to
explore the consequences of breaking spacetime
supersymmetry directly at the string scale.
A generic feature of non--supersymmetric string
vacua is the existence of tachyonic states
in the physical spectrum. Non--supersymmetric
string vacua, such as the $SO(16)\times SO(16)$
heterotic--string in ten dimensions,
do not contain tachyonic states,
but are typically connected in the moduli space
to supersymmetric vacua, and tend to have
large moduli states and group factors. More
realistic constructions on the other hand,
typically have reduced moduli spaces and
contain more sectors that may a priori give rise to tachyons.

It is therefore important to examine the
structure of non--supersymmetric string
vacua in more realistic setting.
In this paper we undertook this task.
We have shown that while generically the
quasi--realistic non--supersymmetric vacua do
contain tachyons, there also exist examples in
which all the tachyonic states are projected out
by the GGSO projections. Furthermore, given that the
moduli spaces of the quasi--realistic constructions
may be much reduced \cite{moduli, cleaver}, one may
entertain the possibility that the tachyon free non--supersymmetric
quasi--realistic vacua may not be connected to supersymmetric
solutions.

We have shown with a concrete example that the
non--supersymmetric quasi--realistic vacuum may retain
some of the structure of the corresponding supersymmetric
solution. This demonstrates that even though supersymmetry
may be broken directly at the string level, the
effective spectrum of the string vacuum, as well as
its low energy effective field theory, may still exhibit
properties that are similar to those of the corresponding
supersymmetric solutions, {\it e.g.} the existence of scalar
replications of the chiral generations.

Another interesting point to note is the existence of an anomalous
$U(1)$ symmetry in the model. The anomalous $U(1)$
is cancelled by the Green--Schwarz--Dine--Seiberg--Witten mechanism
\cite{gs,dsw},
but gives rise to a tadpole diagram
at one--loop order in string perturbation theory \cite{ads},
which reflects the instability of the string vacuum. The
mismatch between the fermionic and bosonic states at different
mass levels gives rise to a non--vanishing vacuum energy, which
similarly gives rise to a tadpole diagram, indicating the
instability of the string vacuum. We may contemplate the possibility
of employing one against the other so that they conspire to
cancel. The anomalous $U(1)$ contribution is proportional
to the trace over the massless fermionic states
and the sign can be altered by the GGSO projections \cite{ads,cfua1}.
It is proportional to the gauge coupling and consequently
only depends on the dilaton moduli. On the other hand, the vacuum amplitude
contribution depends on other moduli \cite{abeldienes}, and
may be tuned to obtain cancellation of the two contributions.
In general, other background fields will be affected by the
shift of the vacuum, and to demonstrate the existence of
a stable vacuum one would need to solve the set of equations
affecting those fields in the shifted vacuum. However,
in this regard the same constraints would apply in the case
of the supersymmetric vacua, where the Fayet--Iliopoulos
term \cite{fi,dsw}, which is generated from the anomalous
$U(1)$ tadpole diagram \cite{dsw,ads},  is cancelled by
assigning VEVs to some massless scalar fields, along flat
supersymmetric directions. It would appear therefore that this
shift of the vacuum is either legitimate, or illegitimate,
in both cases. We therefore propose that the
non--supersymmetric non--tachyonic string vacua should
be considered on equal footing to the supersymmetric
examples.


\section{Acknowledgments}

\noindent
We would like to thank Steve Abel, Ignatios Antoniadis, Costas Kounnas,
Thomas Mohaupt,
Marios Petropoulos and John Rizos for discussions.
A.E.F. would like to thank the Institute for Advanced Study for
hospitality.
A.E.F. is supported in part by STFC under contract  ST/L000431/1.
P.A. acknowledges support from the Hellenic State Scholarship Foundation (IKY).

\appendix

\section{The spectrum of the model in section
\ref{explicitmodel}}\label{sec:Sectors}

\begin{table}[!ht]
\hspace*{-22 mm}
\begin{tabular}{|c|c|c|ccccccc|c|cccc|}
\hline
$F$&
SEC&$(C;L;R)$&$Q_{C}$&$Q_{\bar{\eta}^1}$&$Q_{\bar{\eta}^2}$&$Q_{\bar{\eta}^3}$&$Q_{\bar{y}^{3,6}}$&$Q_{\bar{y}^1\bar{w}^5}$&$Q_{\bar{w}^{2,4}}$&$SU(3)_{H_{1,2}}$&$Q_{\bar{\Phi}^1}$&$Q_{8}$&$Q_{9}$&$Q_{\bar{\Phi}^8}$\\
\hline
b&NS&$(1,1,1)$&\,\,$0$&\,\,$0$&\,\,$0$&\,\,$0$&\,\,$0$&\,\,$0$&\,\,$0$&$(1,1)$&\,\,0&\,\,0&\,\,0&\,\,0\\
&&$(1,1,1)$&\,\,$0$&\,\,$0$&\,\,$0$&\,\,$0$&\,\,$0$&\,\,$0$&\,\,$0$&$(1,1)$&\,\,0&\,\,0&\,\,0&\,\,0\\
&&$(1,1,1)$&\,\,$0$&\,\,$0$&\,\,$0$&\,\,$0$&\,\,$0$&\,\,$0$&\,\,$0$&$(1,1)$&\,\,0&\,\,0&\,\,0&\,\,0\\
&&$(1,1,1)$&\,\,$0$&-$4$&\,\,$4$&\,\,$0$&\,\,$0$&\,\,$0$&\,\,$0$&$(1,1)$&\,\,0&\,\,0&\,\,0&\,\,0\\
&&$(1,1,1)$&\,\,$0$&\,\,$4$&-$4$&\,\,$0$&\,\,$0$&\,\,$0$&\,\,$0$&$(1,1)$&\,\,0&\,\,0&\,\,0&\,\,0\\
&&$(1,1,1)$&\,\,$0$&\,\,$0$&-$4$&\,\,$4$&\,\,$0$&\,\,$0$&\,\,$0$&$(1,1)$&\,\,0&\,\,0&\,\,0&\,\,0\\
&&$(1,1,1)$&\,\,$0$&\,\,$0$&\,\,$4$&-$4$&\,\,$0$&\,\,$0$&\,\,$0$&$(1,1)$&\,\,0&\,\,0&\,\,0&\,\,0\\
&&$(1,1,1)$&\,\,$0$&-$4$&\,\,$0$&\,\,$4$&\,\,$0$&\,\,$0$&\,\,$0$&$(1,1)$&\,\,0&\,\,0&\,\,0&\,\,0\\
&&$(1,1,1)$&\,\,$0$&\,\,$4$&\,\,$0$&-$4$&\,\,$0$&\,\,$0$&\,\,$0$&$(1,1)$&\,\,0&\,\,0&\,\,0&\,\,0\\
\hline

\hline
\end{tabular}
\caption{The untwisted Neveu-Schwarz sector matter states and charges.}
\end{table}

\begin{table}
\hspace*{-22 mm}
\begin{tabular}{|c|c|c|ccccccc|c|cccc|}
\hline
$F$&
SEC&$(C;L;R)$&$Q_{C}$&$Q_{\bar{\eta}^1}$&$Q_{\bar{\eta}^2}$&$Q_{\bar{\eta}^3}$&$Q_{\bar{y}^{3,6}}$&$Q_{\bar{y}^1\bar{w}^5}$&$Q_{\bar{w}^{2,4}}$&$SU(3)_{H_{1,2}}$&$Q_{\bar{\Phi}^1}$&$Q_{8}$&$Q_{9}$&$Q_{\bar{\Phi}^8}$\\
\hline

f&${S}$&$(1,1,1)$&\,\,$0$&\,\,$0$&\,\,$0$&\,\,$0$&\,\,$0$&\,\,$0$&\,\,$0$&$(3,\overline{3})$&\,\,0&-$4$&\,\,$4$&\,\,0\\
&&$(1,1,1)$&\,\,$0$&\,\,$0$&\,\,$0$&\,\,$0$&\,\,$0$&\,\,$0$&\,\,$0$&$(3,\overline{3})$&\,\,0&\,\,4&-$4$&\,\,0\\
&&$(1,1,1)$&\,\,$0$&\,\,$0$&\,\,$0$&\,\,$0$&\,\,$0$&\,\,$0$&\,\,$0$&$(1,1)$&\,\,$4$&\,\,0&\,\,0&\,\,$4$\\
&&$(1,1,1)$&\,\,$0$&\,\,$0$&\,\,$0$&\,\,$0$&\,\,$0$&\,\,$0$&\,\,$0$&$(1,1)$&\,\,$4$&\,\,0&\,\,0&-$4$\\
&&$(1,1,1)$&\,\,$0$&\,\,$0$&\,\,$0$&\,\,$0$&\,\,$0$&\,\,$0$&\,\,$0$&$(1,1)$&-$4$&\,\,0&\,\,0&\,\,$4$\\
&&$(1,1,1)$&\,\,$0$&\,\,$0$&\,\,$0$&\,\,$0$&\,\,$0$&\,\,$0$&\,\,$0$&$(1,1)$&-$4$&\,\,0&\,\,0&-$4$\\

&&$(3,1,1)$       &-$4$&\,\,$4$&\,\,$0$&\,\,$0$&\,\,$0$&\,\,$0$&\,\,$0$&$(1,1)$&\,\,0&\,\,0&\,\,0&\,\,0\\
&&$({\bar 3},1,1)$&\,\,$4$&\,-$4$&\,\,$0$&\,\,$0$&\,\,$0$&\,\,$0$&\,\,$0$&$(1,1)$&\,\,0&\,\,0&\,\,0&\,\,0\\

&&$(3,1,1)$       &-$4$&\,\,$0$&\,\,$4$&\,\,$0$&\,\,$0$&\,\,$0$&\,\,$0$&$(1,1)$&\,\,0&\,\,0&\,\,0&\,\,0\\
&&$({\bar 3},1,1)$&\,\,$4$&\,\,$0$&\,-$4$&\,\,$0$&\,\,$0$&\,\,$0$&\,\,$0$&$(1,1)$&\,\,0&\,\,0&\,\,0&\,\,0\\
&&$(3,1,1)$       &-$4$&\,\,$0$&\,\,$0$&\,\,$4$&\,\,$0$&\,\,$0$&\,\,$0$&$(1,1)$&\,\,0&\,\,0&\,\,0&\,\,0\\
&&$({\bar 3},1,1)$&\,\,$4$&\,\,$0$&\,\,$0$&-$4$&\,\,$0$&\,\,$0$&\,\,$0$&$(1,1)$&\,\,0&\,\,0&\,\,0&\,\,0\\
&&$(1,2,2)$&\,\,$0$&\,\,$0$&\,\,$0$&\,\,$0$&\,\,$0$&\,\,$0$&\,\,$0$&$(1,1)$&\,\,0&\,\,0&\,\,0&\,\,0\\
&&$(1,2,2)$&\,\,$0$&\,\,$0$&\,\,$0$&\,\,$0$&\,\,$0$&\,\,$0$&\,\,$0$&$(1,1)$&\,\,0&\,\,0&\,\,0&\,\,0\\
&&$(1,1,1)$&\,\,$0$&\,\,$0$&\,\,$0$&\,\,$0$&\,\,$4$&\,\,$0$&\,\,$0$&$(1,1)$&\,\,0&\,\,0&\,\,0&\,\,0\\
&&$(1,1,1)$&\,\,$0$&\,\,$0$&\,\,$0$&\,\,$0$&-$4$&\,\,$0$&\,\,$0$&$(1,1)$&\,\,0&\,\,0&\,\,0&\,\,0\\
&&$(1,1,1)$&\,\,$0$&\,\,$0$&\,\,$0$&\,\,$0$&\,\,$0$&\,\,$4$&\,\,$0$&$(1,1)$&\,\,0&\,\,0&\,\,0&\,\,0\\
&&$(1,1,1)$&\,\,$0$&\,\,$0$&\,\,$0$&\,\,$0$&\,\,$0$&-$4$&\,\,$0$&$(1,1)$&\,\,0&\,\,0&\,\,0&\,\,0\\
&&$(1,1,1)$&\,\,$0$&\,\,$0$&\,\,$0$&\,\,$0$&\,\,$4$&\,\,$0$&\,\,$0$&$(1,1)$&\,\,0&\,\,0&\,\,0&\,\,0\\
&&$(1,1,1)$&\,\,$0$&\,\,$0$&\,\,$0$&\,\,$0$&-$4$&\,\,$0$&\,\,$0$&$(1,1)$&\,\,0&\,\,0&\,\,0&\,\,0\\
&&$(1,1,1)$&\,\,$0$&\,\,$0$&\,\,$0$&\,\,$0$&\,\,$0$&\,\,$4$&\,\,$0$&$(1,1)$&\,\,0&\,\,0&\,\,0&\,\,0\\
&&$(1,1,1)$&\,\,$0$&\,\,$0$&\,\,$0$&\,\,$0$&\,\,$0$&-$4$&\,\,$0$&$(1,1)$&\,\,0&\,\,0&\,\,0&\,\,0\\
\hline
\end{tabular}
\caption{The untwisted $S$--sector matter states and charges.}
\end{table}

\newpage
\begin{center}
\begin{table}
\hspace*{-22 mm}
\begin{tabular}{|c|c|c|ccccccc|c|cccc|}
\hline
$F$&
SEC&$(C;L;R)$&$Q_{C}$&$Q_{\bar{\eta}^1}$&$Q_{\bar{\eta}^2}$&$Q_{\bar{\eta}^3}$&$Q_{\bar{y}^{3,6}}$&$Q_{\bar{y}^1\bar{w}^5}$&$Q_{\bar{w}^{2,4}}$&$SU(3)_{H_{1,2}}$&$Q_{\bar{\Phi}^1}$&$Q_{8}$&$Q_{9}$&$Q_{\bar{\Phi}^8}$\\
\hline
$Q_{L_{1}}$&${{b}}_{1}$&$(3,2,1)$&\,\,$2$&\,\,$2$&\,\,$0$&\,\,$0$&-$2$&\,\,$0$&\,\,$0$&$(1,1)$&0&0&0&0\\
$Q_{R_{1}}$&&$(\overline{3},1,2)$&-$2$&-$2$&\,\,$0$&\,\,$0$&-$2$&\,\,$0$&\,\,$0$&$(1,1)$&0&0&0&0 \\
$L_{L_{1}}$&&$(1,2,1)$&-$6$&\,\,$2$&\,\,$0$&\,\,$0$&-$2$&\,\,$0$&\,\,$0$&$(1,1)$&0&0&0&0\\
$L_{R_{1}}$&&$(1,1,2)$&\,\,$6$&-$2$&\,\,$0$&\,\,$0$&-$2$&\,\,$0$&\,\,$0$&$(1,1)$&0&0&0&0\\
\hline
b&${S}+{b_{1}}$&$(3,1,2)$&\,\,$2$&\,\,$2$&\,\,$0$&\,\,$0$&-$2$&\,\,$0$&\,\,$0$&$(1,1)$&$0$&$0$&$0$&$0$\\
&&$(\overline{3},2,1)$&-$2$&-$2$&\,\,$0$&\,\,$0$&-$2$&\,\,$0$&\,\,$0$&$(1,1)$&$0$&$0$&$0$&$0$\\
&&$(1,2,1)$&\,\,$6$&-$2$&\,\,$0$&\,\,$0$&-$2$&\,\,$0$&\,\,$0$&$(1,1)$&$0$&$0$&$0$&$0$\\
&&$(1,1,2)$&-$6$&\,\,$2$&\,\,$0$&\,\,$0$&-$2$&\,\,$0$&\,\,$0$&$(1,1)$&$0$&$0$&$0$&$0$\\
\hline
$Q_{L_{2}}$&${{b}}_{2}$&$(3,2,1)$&\,\,$2$&\,\,$0$&\,\,$2$&\,\,$0$&\,\,$0$&-$2$&\,\,$0$&$(1,1)$&0&0&0&0\\
$Q_{R_{2}}$&&$(\overline{3},1,2)$&-$2$&\,\,$0$&-$2$&\,\,$0$&\,\,$0$&-$2$&\,\,$0$&$(1,1)$&0&0&0&0 \\
$L_{L_{2}}$&&$(1,2,1)$&-$6$&\,\,$0$&\,\,$2$&\,\,$0$&\,\,$0$&-$2$&\,\,$0$&$(1,1)$&0&0&0&0\\
$L_{R_{2}}$&&$(1,1,2)$&\,\,$6$&\,\,$0$&-$2$&\,\,$0$&\,\,$0$&-$2$&\,\,$0$&$(1,1)$&0&0&0&0\\
\hline
b&${S}+{b_{2}}$&$(3,1,2)$&\,\,$2$&\,\,$0$&\,\,$2$&\,\,$0$&\,\,$0$&-$2$&\,\,$0$&$(1,1)$&$0$&$0$&$0$&$0$\\
&&$(\overline{3},2,1)$&-$2$&\,\,$0$&-$2$&\,\,$0$&\,\,$0$&-$2$&\,\,$0$&$(1,1)$&$0$&$0$&$0$&$0$\\
&&$(1,2,1)$&\,\,$6$&\,\,$0$&-$2$&\,\,$0$&\,\,$0$&-$2$&\,\,$0$&$(1,1)$&$0$&$0$&$0$&$0$\\
&&$(1,1,2)$&-$6$&\,\,$0$&\,\,$2$&\,\,$0$&\,\,$0$&-$2$&\,\,$0$&$(1,1)$&$0$&$0$&$0$&$0$\\
\hline
$Q_{L_{3}}$&${{b}}_{3}$&$(3,2,1)$&\,\,$2$&\,\,$0$&\,\,$0$&\,\,$2$&\,\,$0$&\,\,$0$&-$2$&$(1,1)$&0&0&0&0\\
$Q_{R_{3}}$&&$(\overline{3},1,2)$&-$2$&\,\,$0$&\,\,$0$&-$2$&\,\,$0$&\,\,$0$&-$2$&$(1,1)$&0&0&0&0 \\
$L_{L_{3}}$&&$(1,2,1)$&-$6$&\,\,$0$&\,\,$0$&\,\,$2$&\,\,$0$&\,\,$0$&-$2$&$(1,1)$&0&0&0&0\\
$L_{R_{3}}$&&$(1,1,2)$&\,\,$6$&\,\,$0$&\,\,$0$&-$2$&\,\,$0$&\,\,$0$&-$2$&$(1,1)$&0&0&0&0\\
\hline
b&${S}+{b_{3}}$&$(3,1,2)$&\,\,$2$&\,\,$0$&\,\,$0$&\,\,$2$&\,\,$0$&\,\,$0$&-$2$&$(1,1)$&$0$&$0$&$0$&$0$\\
&&$(\overline{3},2,1)$&-$2$&\,\,$0$&\,\,$0$&-$2$&\,\,$0$&\,\,$0$&-$2$&$(1,1)$&$0$&$0$&$0$&$0$\\
&&$(1,2,1)$&\,\,$6$&\,\,$0$&\,\,$0$&-$2$&\,\,$0$&\,\,$0$&-$2$&$(1,1)$&$0$&$0$&$0$&$0$\\
&&$(1,1,2)$&-$6$&\,\,$0$&\,\,$0$&\,\,$2$&\,\,$0$&\,\,$0$&-$2$&$(1,1)$&$0$&$0$&$0$&$0$\\
\hline
\end{tabular}
\caption{The observable matter sectors. All sectors, fermionic and bosonic, have CPT conjugates which are not displayed.}
\end{table}
\end{center}

\newpage
\begin{center}
\begin{table}[!ht]
\hspace*{-22 mm}
\begin{tabular}{|c|c|c|ccccccc|c|cccc|}
\hline
$F$&
SEC&$(C;L;R)$&$Q_{C}$&$Q_{\bar{\eta}^1}$&$Q_{\bar{\eta}^2}$&$Q_{\bar{\eta}^3}$&$Q_{\bar{y}^{3,6}}$&$Q_{\bar{y}^1\bar{w}^5}$&$Q_{\bar{w}^{2,4}}$&$SU(3)_{H_{1,2}}$&$Q_{\bar{\Phi}^1}$&$Q_{8}$&$Q_{9}$&$Q_{\bar{\Phi}^8}$\\
\hline
f&${S}+$&$(1,1,1)$&\,\,$0$&\,\,$2$&-$2$&\,\,$0$&\,\,$0$&\,\,$0$&\,\,$0$&$(1,1)$&\,\,$0$&\,\,$0$&\,\,$0$&\,\,$4$\\
&${b_{1}}+{b_{2}}$&$(1,1,1)$&\,\,$0$&\,\,$2$&-$2$&\,\,$0$&\,\,$0$&\,\,$0$&\,\,$0$&$(1,1)$&\,\,$0$&\,\,$0$&\,\,$0$&-$4$\\
&$+\alpha+\beta$&$(1,1,1)$&\,\,$0$&-$2$&\,\,$2$&\,\,$0$&\,\,$0$&\,\,$0$&\,\,$0$&$(1,1)$&\,\,$0$&\,\,$0$&\,\,$0$&\,\,$4$\\
&&$(1,1,1)$&\,\,$0$&-$2$&\,\,$2$&\,\,$0$&\,\,$0$&\,\,$0$&\,\,$0$&$(1,1)$&\,\,$0$&\,\,$0$&\,\,$0$&-$4$\\
&&$(1,1,1)$&\,\,$0$&\,\,$2$&\,\,$2$&\,\,$0$&\,\,$0$&\,\,$0$&\,\,$0$&$(3,1)$&\,\,$0$&\,\,$4$&\,\,$0$&\,\,$0$\\
&&$(1,1,1)$&\,\,$0$&-$2$&-$2$&\,\,$0$&\,\,$0$&\,\,$0$&\,\,$0$&$(3,1)$&\,\,$0$&\,\,$4$&\,\,$0$&\,\,$0$\\
&&$(1,1,1)$&\,\,$0$&\,\,$2$&\,\,$2$&\,\,$0$&\,\,$0$&\,\,$0$&\,\,$0$&$(\overline{3},1)$&\,\,$0$&-$4$&\,\,$0$&\,\,$0$\\
&&$(1,1,1)$&\,\,$0$&-$2$&-$2$&\,\,$0$&\,\,$0$&\,\,$0$&\,\,$0$&$(\overline{3},1)$&\,\,$0$&-$4$&\,\,$0$&\,\,$0$\\
\hline
b&${b_{1}}+{b_{2}}$&$(1,1,1)$&\,\,$0$&\,\,$2$&-$2$&\,\,$0$&\,\,$0$&\,\,$0$&\,\,$0$&$(1,1)$&\,\,$4$&\,\,$0$&\,\,$0$&\,\,$0$\\
&$+\alpha+\beta$&$(1,1,1)$&\,\,$0$&\,\,$2$&-$2$&\,\,$0$&\,\,$0$&\,\,$0$&\,\,$0$&$(1,1)$&-$4$&\,\,$0$&\,\,$0$&\,\,$0$\\
&&$(1,1,1)$&\,\,$0$&-$2$&\,\,$2$&\,\,$0$&\,\,$0$&\,\,$0$&\,\,$0$&$(1,1)$&\,\,$4$&\,\,$0$&\,\,$0$&\,\,$0$\\
&&$(1,1,1)$&\,\,$0$&-$2$&\,\,$2$&\,\,$0$&\,\,$0$&\,\,$0$&\,\,$0$&$(1,1)$&-$4$&\,\,$0$&\,\,$0$&\,\,$0$\\
&&$(1,1,1)$&\,\,$0$&\,\,$2$&\,\,$2$&\,\,$0$&\,\,$0$&\,\,$0$&\,\,$0$&$(1,3)$&\,\,$0$&\,\,$0$&\,\,$4$&\,\,$0$\\
&&$(1,1,1)$&\,\,$0$&-$2$&-$2$&\,\,$0$&\,\,$0$&\,\,$0$&\,\,$0$&$(1,3)$&\,\,$0$&\,\,$0$&\,\,$4$&\,\,$0$\\
&&$(1,1,1)$&\,\,$0$&\,\,$2$&\,\,$2$&\,\,$0$&\,\,$0$&\,\,$0$&\,\,$0$&$(1,\overline{3})$&\,\,$0$&\,\,$0$&-$4$&\,\,$0$\\
&&$(1,1,1)$&\,\,$0$&-$2$&-$2$&\,\,$0$&\,\,$0$&\,\,$0$&\,\,$0$&$(1,\overline{3})$&\,\,$0$&\,\,$0$&-$4$&\,\,$0$\\
\hline
\end{tabular}
\caption{Vector-like $SO(10)$ singlet states. All sectors, fermionic and bosonic, have CPT conjugates which are not displayed.}
\end{table}
\end{center}

\newpage
\begin{center}
\begin{table}[!ht]
\hspace*{-22 mm}
\begin{tabular}{|c|c|c|ccccccc|c|cccc|}
\hline
$F$&
SEC&$(C;L;R)$&$Q_{C}$&$Q_{\bar{\eta}^1}$&$Q_{\bar{\eta}^2}$&$Q_{\bar{\eta}^3}$&$Q_{\bar{y}^{3,6}}$&$Q_{\bar{y}^1\bar{w}^5}$&$Q_{\bar{w}^{2,4}}$&$SU(3)_{H_{1,2}}$&$Q_{\bar{\Phi}^1}$&$Q_{8}$&$Q_{9}$&$Q_{\bar{\Phi}^8}$\\
\hline
f&${S}+$&$(1,1,1)$&\,\,$0$&-$2$&\,\,$0$&\,\,$0$&\,\,$0$&-$2$&-$2$&$(1,1)$&$0$&\,\,$6$&$0$&\,\,$2$\\
&${1}+{b_{1}}$&$(1,1,1)$&\,\,$0$&\,\,$2$&\,\,$0$&\,\,$0$&\,\,$0$&-$2$&-$2$&$(1,1)$&$0$&-$6$&$0$&-$2$\\
&$+\alpha+2\gamma$&$(1,1,1)$&\,\,$0$&\,\,$2$&\,\,$0$&\,\,$0$&\,\,$0$&-$2$&-$2$&$(3,1)$&$0$&\,\,$2$&$0$&-$2$\\
&&$(1,1,1)$&\,\,$0$&-$2$&\,\,$0$&\,\,$0$&\,\,$0$&-$2$&-$2$&$(\overline{3},1)$&$0$&-$2$&$0$&\,\,$2$\\
\hline
b&${1}+{b_{1}}$&$(1,1,1)$&\,\,$0$&-$2$&\,\,$0$&\,\,$0$&\,\,$0$&\,\,$2$&-$2$&$(1,1)$&\,\,$0$&\,\,$6$&\,\,$0$&\,\,$2$\\
&$+\alpha+2\gamma$&$(1,1,1)$&\,\,$0$&\,\,$2$&\,\,$0$&\,\,$0$&\,\,$0$&\,\,$2$&-$2$&$(1,1)$&\,\,$0$&-$6$&\,\,$0$&-$2$\\
&&$(1,1,1)$&\,\,$0$&\,\,$2$&\,\,$0$&\,\,$0$&\,\,$0$&\,\,$2$&-$2$&$(3,1)$&\,\,$0$&\,\,$2$&\,\,$0$&-$2$\\
&&$(1,1,1)$&\,\,$0$&-$2$&\,\,$0$&\,\,$0$&\,\,$0$&\,\,$2$&-$2$&$(\overline{3},1)$&\,\,$0$&-$2$&\,\,$0$&\,\,$2$\\
\hline
f&${S}+$&$(1,1,1)$&\,\,$0$&\,\,$0$&-$2$&\,\,$0$&-$2$&\,\,$0$&-$2$&$(1,1)$&\,\,$0$&\,\,$6$&\,\,$0$&\,\,$2$\\
&${1}+{b_{2}}$&$(1,1,1)$&\,\,$0$&\,\,$0$&\,\,$2$&\,\,$0$&-$2$&\,\,$0$&-$2$&$(1,1)$&\,\,$0$&-$6$&\,\,$0$&-$2$\\
&$+\alpha+2\gamma$&$(1,1,1)$&\,\,$0$&\,\,$0$&\,\,$2$&\,\,$0$&-$2$&\,\,$0$&-$2$&$(3,1)$&\,\,$0$&\,\,$2$&\,\,$0$&-$2$\\
&&$(1,1,1)$&\,\,$0$&\,\,$0$&-$2$&\,\,$0$&-$2$&\,\,$0$&-$2$&$(\overline{3},1)$&\,\,$0$&-$2$&\,\,$0$&\,\,$2$\\
\hline
b&${1}+{b_{2}}$&$(1,1,1)$&\,\,$0$&\,\,$0$&-$2$&\,\,$0$&\,\,$2$&\,\,$0$&-$2$&$(1,1)$&\,\,$0$&\,\,$6$&\,\,$0$&\,\,$2$\\
&$+\alpha+2\gamma$&$(1,1,1)$&\,\,$0$&\,\,$0$&\,\,$2$&\,\,$0$&\,\,$2$&\,\,$0$&-$2$&$(1,1)$&\,\,$0$&-$6$&\,\,$0$&-$2$\\
&&$(1,1,1)$&\,\,$0$&\,\,$0$&\,\,$2$&\,\,$0$&\,\,$2$&\,\,$0$&-$2$&$(3,1)$&\,\,$0$&\,\,$2$&\,\,$0$&-$2$\\
&&$(1,1,1)$&\,\,$0$&\,\,$0$&-$2$&\,\,$0$&\,\,$2$&\,\,$0$&-$2$&$(\overline{3},1)$&\,\,$0$&-$2$&\,\,$0$&\,\,$2$\\
\hline
f&${S}+$&$(1,1,1)$&\,\,$0$&\,\,$0$&-$2$&\,\,$0$&-$2$&\,\,$0$&-$2$&$(1,1)$&-$2$&\,\,$0$&\,\,$6$&\,\,$0$\\
&${b_{1}}+{b_{3}}$&$(1,1,1)$&\,\,$0$&\,\,$0$&\,\,$2$&\,\,$0$&-$2$&\,\,$0$&-$2$&$(1,1)$&\,\,$2$&\,\,$0$&-$6$&\,\,$0$\\
&$+\alpha+2\gamma$&$(1,1,1)$&\,\,$0$&\,\,$0$&\,\,$2$&\,\,$0$&-$2$&\,\,$0$&-$2$&$(1,3)$&\,\,$2$&\,\,$0$&\,\,$2$&\,\,$0$\\
&&$(1,1,1)$&\,\,$0$&\,\,$0$&-$2$&\,\,$0$&-$2$&\,\,$0$&-$2$&$(1,\overline{3})$&-$2$&\,\,$0$&-$2$&\,\,$0$\\
\hline
b&${b_{1}}+{b_{3}}$&$(1,1,1)$&\,\,$0$&\,\,$0$&-$2$&\,\,$0$&\,\,$2$&\,\,$0$&-$2$&$(1,1)$&-$2$&\,\,$0$&\,\,$6$&\,\,$0$\\
&$+\alpha+2\gamma$&$(1,1,1)$&\,\,$0$&\,\,$0$&\,\,$2$&\,\,$0$&\,\,$2$&\,\,$0$&-$2$&$(1,1)$&\,\,$2$&\,\,$0$&-$6$&\,\,$0$\\
&&$(1,1,1)$&\,\,$0$&\,\,$0$&\,\,$2$&\,\,$0$&\,\,$2$&\,\,$0$&-$2$&$(1,3)$&\,\,$2$&\,\,$0$&\,\,$2$&\,\,$0$\\
&&$(1,1,1)$&\,\,$0$&\,\,$0$&-$2$&\,\,$0$&\,\,$2$&\,\,$0$&-$2$&$(1,\overline{3})$&-$2$&\,\,$0$&-$2$&\,\,$0$\\
\hline
f&${S}+$&$(1,1,1)$&\,\,$0$&\,\,$0$&\,\,$0$&-$2$&-$2$&-$2$&\,\,$0$&$(1,1)$&-$2$&\,\,$0$&\,\,$6$&\,\,$0$\\
&${b_{1}}+{b_{2}}$&$(1,1,1)$&\,\,$0$&\,\,$0$&\,\,$0$&\,\,$2$&-$2$&-$2$&\,\,$0$&$(1,1)$&\,\,$2$&\,\,$0$&-$6$&\,\,$0$\\
&$+\alpha+2\gamma$&$(1,1,1)$&\,\,$0$&\,\,$0$&\,\,$0$&\,\,$2$&-$2$&-$2$&\,\,$0$&$(1,3)$&\,\,$2$&\,\,$0$&\,\,$2$&\,\,$0$\\
&&$(1,1,1)$&\,\,$0$&\,\,$0$&\,\,$0$&-$2$&-$2$&-$2$&\,\,$0$&$(1,\overline{3})$&-$2$&\,\,$0$&-$2$&\,\,$0$\\
\hline
b&${b_{1}}+{b_{2}}$&$(1,1,1)$&\,\,$0$&\,\,$0$&\,\,$0$&-$2$&\,\,$2$&-$2$&\,\,$0$&$(1,1)$&-$2$&\,\,$0$&\,\,$6$&\,\,$0$\\
&$+\alpha+2\gamma$&$(1,1,1)$&\,\,$0$&\,\,$0$&\,\,$0$&\,\,$2$&\,\,$2$&-$2$&\,\,$0$&$(1,1)$&\,\,$2$&\,\,$0$&-$6$&\,\,$0$\\
&&$(1,1,1)$&\,\,$0$&\,\,$0$&\,\,$0$&\,\,$2$&\,\,$2$&-$2$&\,\,$0$&$(1,3)$&\,\,$2$&\,\,$0$&\,\,$2$&\,\,$0$\\
&&$(1,1,1)$&\,\,$0$&\,\,$0$&\,\,$0$&-$2$&\,\,$2$&-$2$&\,\,$0$&$(1,\overline{3})$&-$2$&\,\,$0$&-$2$&\,\,$0$\\
\hline
\end{tabular}
\caption{Vector-like $SO(10)$ singlet states. All sectors, fermionic and bosonic, have CPT conjugates which are not displayed.}
\end{table}
\end{center}

\newpage
\begin{center}
\begin{table}[!ht]
\hspace*{-22 mm}
\begin{tabular}{|c|c|c|ccccccc|c|cccc|}
\hline
$F$&
SEC&$(C;L;R)$&$Q_{C}$&$Q_{\bar{\eta}^1}$&$Q_{\bar{\eta}^2}$&$Q_{\bar{\eta}^3}$&$Q_{\bar{y}^{3,6}}$&$Q_{\bar{y}^1\bar{w}^5}$&$Q_{\bar{w}^{2,4}}$&$SU(3)_{H_{1,2}}$&$Q_{\bar{\Phi}^1}$&$Q_{8}$&$Q_{9}$&$Q_{\bar{\Phi}^8}$\\
\hline
f&${S}+$&$(1,1,1)$&\,\,$0$&-$2$&\,\,$0$&\,\,$0$&\,\,$0$&-$2$&-$2$&$(1,1)$&-$2$&\,\,$0$&\,\,$6$&\,\,$0$\\
&${b_{2}}+{b_{3}}$&$(1,1,1)$&\,\,$0$&\,\,$2$&\,\,$0$&\,\,$0$&\,\,$0$&-$2$&-$2$&$(1,1)$&\,\,$2$&\,\,$0$&-$6$&\,\,$0$\\
&$+\alpha+2\gamma$&$(1,1,1)$&\,\,$0$&\,\,$2$&\,\,$0$&\,\,$0$&\,\,$0$&-$2$&-$2$&$(1,3)$&\,\,$2$&\,\,$0$&\,\,$2$&\,\,$0$\\
&&$(1,1,1)$&\,\,$0$&-$2$&\,\,$0$&\,\,$0$&\,\,$0$&-$2$&-$2$&$(1,\overline{3})$&-$2$&\,\,$0$&-$2$&\,\,$0$\\
\hline
b&${b_{2}}+{b_{3}}$&$(1,1,1)$&\,\,$0$&-$2$&\,\,$0$&\,\,$0$&\,\,$0$&\,\,$2$&-$2$&$(1,1)$&-$2$&\,\,$0$&\,\,$6$&\,\,$0$\\
&$+\alpha+2\gamma$&$(1,1,1)$&\,\,$0$&\,\,$2$&\,\,$0$&\,\,$0$&\,\,$0$&\,\,$2$&-$2$&$(1,1)$&\,\,$2$&$0$&-$6$&\,\,$0$\\
&&$(1,1,1)$&\,\,$0$&\,\,$2$&\,\,$0$&\,\,$0$&\,\,$0$&\,\,$2$&-$2$&$(1,3)$&\,\,$2$&$0$&\,\,$2$&\,\,$0$\\
&&$(1,1,1)$&\,\,$0$&-$2$&\,\,$0$&\,\,$0$&\,\,$0$&\,\,$2$&-$2$&$(1,\overline{3})$&-$2$&$0$&-$2$&\,\,$0$\\
\hline
f&${S}+$&$(1,1,1)$&\,\,$0$&\,\,$0$&\,\,$0$&-$2$&-$2$&-$2$&\,\,$0$&$(1,1)$&\,\,$0$&\,\,$6$&\,\,$0$&\,\,$2$\\
&${1}+{b_{3}}$&$(1,1,1)$&\,\,$0$&\,\,$0$&\,\,$0$&\,\,$2$&-$2$&-$2$&\,\,$0$&$(1,1)$&\,\,$0$&-$6$&\,\,$0$&-$2$\\
&$+\alpha+2\gamma$&$(1,1,1)$&\,\,$0$&\,\,$0$&\,\,$0$&\,\,$2$&-$2$&-$2$&\,\,$0$&$(3,1)$&\,\,$0$&\,\,$2$&\,\,$0$&-$2$\\
&&$(1,1,1)$&\,\,$0$&\,\,$0$&\,\,$0$&-$2$&-$2$&-$2$&\,\,$0$&$(\overline{3},1)$&\,\,$0$&-$2$&\,\,$0$&\,\,$2$\\
\hline
b&${1}+{b_{3}}$&$(1,1,1)$&\,\,$0$&\,\,$0$&\,\,$0$&-$2$&\,\,$2$&-$2$&\,\,$0$&$(1,1)$&\,\,$0$&\,\,$6$&\,\,$0$&\,\,$2$\\
&$+\alpha+2\gamma$&$(1,1,1)$&\,\,$0$&\,\,$0$&\,\,$0$&\,\,$2$&\,\,$2$&-$2$&\,\,$0$&$(1,1)$&\,\,$0$&-$6$&\,\,$0$&-$2$\\
&&$(1,1,1)$&\,\,$0$&\,\,$0$&\,\,$0$&\,\,$2$&\,\,$2$&-$2$&\,\,$0$&$(3,1)$&\,\,$0$&\,\,$2$&\,\,$0$&-$2$\\
&&$(1,1,1)$&\,\,$0$&\,\,$0$&\,\,$0$&-$2$&\,\,$2$&-$2$&\,\,$0$&$(\overline{3},1)$&\,\,$0$&-$2$&\,\,$0$&\,\,$2$\\
\hline
\end{tabular}
\caption{Table 5 continued.}
\end{table}
\end{center}

\newpage
\begin{center}
\begin{table}[!ht]
\hspace*{-22 mm}
\begin{tabular}{|c|c|c|ccccccc|c|cccc|}
\hline
$F$&
SEC&$(C;L;R)$&$Q_{C}$&$Q_{\bar{\eta}^1}$&$Q_{\bar{\eta}^2}$&$Q_{\bar{\eta}^3}$&$Q_{\bar{y}^{3,6}}$&$Q_{\bar{y}^1\bar{w}^5}$&$Q_{\bar{w}^{2,4}}$&$SU(3)_{H_{1,2}}$&$Q_{\bar{\Phi}^1}$&$Q_{8}$&$Q_{9}$&$Q_{\bar{\Phi}^8}$\\
\hline
b&$\alpha\pm\gamma$&$(1,1,1)$&-$3$&\,\,$1$&\,\,$1$&\,\,$1$&-$2$&\,\,$0$&-$2$&$(1,1)$&\,\,$2$&-$3$&\,\,$3$&\,\,$0$\\
&&$(1,1,1)$&-$3$&\,\,$1$&\,\,$1$&\,\,$1$&\,\,$2$&\,\,$0$&\,\,$2$&$(1,1)$&\,\,$2$&-$3$&\,\,$3$&\,\,$0$\\
&&$(1,1,1)$&-$3$&\,\,$1$&\,\,$1$&\,\,$1$&\,\,$2$&\,\,$0$&\,\,$2$&$(1,1)$&\,\,$2$&-$3$&\,\,$3$&\,\,$0$\\
&&$(1,1,1)$&-$3$&\,\,$1$&\,\,$1$&\,\,$1$&-$2$&\,\,$0$&-$2$&$(1,1)$&\,\,$2$&-$3$&\,\,$3$&\,\,$0$\\
&&$(1,1,1)$&\,\,$3$&-$1$&-$1$&-$1$&-$2$&\,\,$0$&-$2$&$(1,1)$&-$2$&\,\,$3$&-$3$&\,\,$0$\\
&&$(1,1,1)$&\,\,$3$&-$1$&-$1$&-$1$&\,\,$2$&\,\,$0$&\,\,$2$&$(1,1)$&-$2$&\,\,$3$&-$3$&\,\,$0$\\
&&$(1,1,1)$&\,\,$3$&-$1$&-$1$&-$1$&\,\,$2$&\,\,$0$&\,\,$2$&$(1,1)$&-$2$&\,\,$3$&-$3$&\,\,$0$\\
&&$(1,1,1)$&\,\,$3$&-$1$&-$1$&-$1$&-$2$&\,\,$0$&-$2$&$(1,1)$&-$2$&\,\,$3$&-$3$&\,\,$0$\\
\hline
b&$\beta\pm\gamma$&$(1,1,1)$&-$3$&\,\,$1$&\,\,$1$&\,\,$1$&\,\,$0$&\,\,$2$&\,\,$2$&$(1,1)$&-$2$&-$3$&\,\,$3$&\,\,$0$\\
&&$(1,1,1)$&-$3$&\,\,$1$&\,\,$1$&\,\,$1$&\,\,$0$&-$2$&-$2$&$(1,1)$&-$2$&-$3$&\,\,$3$&\,\,$0$\\
&&$(1,1,1)$&-$3$&\,\,$1$&\,\,$1$&\,\,$1$&\,\,$0$&-$2$&-$2$&$(1,1)$&-$2$&-$3$&\,\,$3$&\,\,$0$\\
&&$(1,1,1)$&-$3$&\,\,$1$&\,\,$1$&\,\,$1$&\,\,$0$&\,\,$2$&\,\,$2$&$(1,1)$&-$2$&-$3$&\,\,$3$&\,\,$0$\\
&&$(1,1,1)$&\,\,$3$&-$1$&-$1$&-$1$&\,\,$0$&\,\,$2$&\,\,$2$&$(1,1)$&\,\,$2$&\,\,$3$&-$3$&\,\,$0$\\
&&$(1,1,1)$&\,\,$3$&-$1$&-$1$&-$1$&\,\,$0$&-$2$&-$2$&$(1,1)$&\,\,$2$&\,\,$3$&-$3$&\,\,$0$\\
&&$(1,1,1)$&\,\,$3$&-$1$&-$1$&-$1$&\,\,$0$&-$2$&-$2$&$(1,1)$&\,\,$2$&\,\,$3$&-$3$&\,\,$0$\\
&&$(1,1,1)$&\,\,$3$&-$1$&-$1$&-$1$&\,\,$0$&\,\,$2$&\,\,$2$&$(1,1)$&\,\,$2$&\,\,$3$&-$3$&\,\,$0$\\
\hline
b&${1}+{b_{1}}$&$(1,1,1)$&-$3$&\,\,$1$&\,\,$1$&\,\,$1$&\,\,$0$&\,\,$2$&\,\,$2$&$(1,1)$&\,\,$0$&\,\,$3$&-$3$&\,\,$2$\\
&$+{{b_{2}}}+{{b_{3}}}$&$(1,1,1)$&-$3$&\,\,$1$&\,\,$1$&\,\,$1$&\,\,$0$&-$2$&-$2$&$(1,1)$&\,\,$0$&\,\,$3$&-$3$&\,\,$2$\\
&$+\beta\pm\gamma$&$(1,1,1)$&-$3$&\,\,$1$&\,\,$1$&\,\,$1$&\,\,$0$&-$2$&-$2$&$(1,1)$&\,\,$0$&\,\,$3$&-$3$&\,\,$2$\\
&&$(1,1,1)$&-$3$&\,\,$1$&\,\,$1$&\,\,$1$&\,\,$0$&\,\,$2$&\,\,$2$&$(1,1)$&\,\,$0$&\,\,$3$&-$3$&\,\,$2$\\
&&$(1,1,1)$&\,\,$3$&-$1$&-$1$&-$1$&\,\,$0$&\,\,$2$&\,\,$2$&$(1,1)$&\,\,$0$&-$3$&\,\,$3$&-$2$\\
&&$(1,1,1)$&\,\,$3$&-$1$&-$1$&-$1$&\,\,$0$&-$2$&-$2$&$(1,1)$&\,\,$0$&-$3$&\,\,$3$&-$2$\\
&&$(1,1,1)$&\,\,$3$&-$1$&-$1$&-$1$&\,\,$0$&-$2$&-$2$&$(1,1)$&\,\,$0$&-$3$&\,\,$3$&-$2$\\
&&$(1,1,1)$&\,\,$3$&-$1$&-$1$&-$1$&\,\,$0$&\,\,$2$&\,\,$2$&$(1,1)$&\,\,$0$&-$3$&\,\,$3$&-$2$\\
\hline
b&${1}+{b_{1}}$&$(1,1,1)$&-$3$&\,\,$1$&\,\,$1$&\,\,$1$&\,\,$2$&\,\,$0$&-$2$&$(1,1)$&\,\,$0$&\,\,$3$&-$3$&-$2$\\
&$+{{b_{2}}}+{{b_{3}}}$&$(1,1,1)$&-$3$&\,\,$1$&\,\,$1$&\,\,$1$&-$2$&\,\,$0$&\,\,$2$&$(1,1)$&\,\,$0$&\,\,$3$&-$3$&-$2$\\
&$+\alpha\pm \gamma$&$(1,1,1)$&-$3$&\,\,$1$&\,\,$1$&\,\,$1$&\,\,$2$&\,\,$0$&-$2$&$(1,1)$&\,\,$0$&\,\,$3$&-$3$&-$2$\\
&&$(1,1,1)$&-$3$&\,\,$1$&\,\,$1$&\,\,$1$&-$2$&\,\,$0$&\,\,$2$&$(1,1)$&\,\,$0$&\,\,$3$&-$3$&-$2$\\
&&$(1,1,1)$&\,\,$3$&-$1$&-$1$&-$1$&\,\,$2$&\,\,$0$&-$2$&$(1,1)$&\,\,$0$&-$3$&\,\,$3$&\,\,$2$\\
&&$(1,1,1)$&\,\,$3$&-$1$&-$1$&-$1$&-$2$&\,\,$0$&\,\,$2$&$(1,1)$&\,\,$0$&-$3$&\,\,$3$&\,\,$2$\\
&&$(1,1,1)$&\,\,$3$&-$1$&-$1$&-$1$&\,\,$2$&\,\,$0$&-$2$&$(1,1)$&\,\,$0$&-$3$&\,\,$3$&\,\,$2$\\
&&$(1,1,1)$&\,\,$3$&-$1$&-$1$&-$1$&-$2$&\,\,$0$&\,\,$2$&$(1,1)$&\,\,$0$&-$3$&\,\,$3$&\,\,$2$\\
\hline
\end{tabular}
\caption{The table displays all the massless sectors for which the ``would-be superpartners" are massive and do not form part of the massless spectrum. The ``would-be superpartners" arise from the sectors that are obtained by adding the basis vector $S$ to a given sector and are the  fermionic counterparts.}
\end{table}
\end{center}

\newpage
\begin{center}
\begin{table}
\hspace*{-22 mm}
\begin{tabular}{|c|c|c|ccccccc|c|cccc|}
\hline
$F$&
SEC&$(C;L;R)$&$Q_{C}$&$Q_{\bar{\eta}^1}$&$Q_{\bar{\eta}^2}$&$Q_{\bar{\eta}^3}$&$Q_{\bar{y}^{3,6}}$&$Q_{\bar{y}^1\bar{w}^5}$&$Q_{\bar{w}^{2,4}}$&$SU(3)_{H_{1,2}}$&$Q_{\bar{\Phi}^1}$&$Q_{8}$&$Q_{9}$&$Q_{\bar{\Phi}^8}$\\
\hline
f&${S}+$&$(1,1,1)$&-$3$&-$3$&-$1$&-$1$&\,\,$0$&\,\,$0$&\,\,$0$&$(1,1)$&-$2$&-$3$&\,\,$3$&\,\,$0$\\
&${b_{2}}+{b_{3}}$&$(1,1,1)$&-$3$&\,\,$1$&\,\,$3$&-$1$&\,\,$0$&\,\,$0$&\,\,$0$&$(1,1)$&-$2$&-$3$&\,\,$3$&\,\,$0$\\
&$+\beta\pm\gamma$&$(1,1,1)$&-$3$&\,\,$1$&-$1$&\,\,$3$&\,\,$0$&\,\,$0$&\,\,$0$&$(1,1)$&-$2$&-$3$&\,\,$3$&\,\,$0$\\
&&$(1,1,1)$&\,\,$3$&-$1$&\,\,$1$&\,\,$1$&\,\,$0$&\,\,$0$&\,\,$0$&$(1,\overline{3})$&-$2$&\,\,$3$&\,\,$1$&\,\,$0$\\
&&$(1,1,1)$&\,\,$3$&\,\,$3$&\,\,$1$&\,\,$1$&\,\,$0$&\,\,$0$&\,\,$0$&$(1,1)$&\,\,$2$&\,\,$3$&-$3$&\,\,$0$\\
&&$(1,1,1)$&\,\,$3$&-$1$&-$3$&\,\,$1$&\,\,$0$&\,\,$0$&\,\,$0$&$(1,1)$&\,\,$2$&\,\,$3$&-$3$&\,\,$0$\\
&&$(1,1,1)$&\,\,$3$&-$1$&\,\,$1$&-$3$&\,\,$0$&\,\,$0$&\,\,$0$&$(1,1)$&\,\,$2$&\,\,$3$&-$3$&\,\,$0$\\
&&$(1,1,1)$&-$3$&\,\,$1$&-$1$&-$1$&\,\,$0$&\,\,$0$&\,\,$0$&$(1,3)$&\,\,$2$&-$3$&-$1$&\,\,$0$\\
\hline
b&${b_{2}}+{b_{3}}$&$(\overline{3},1,1)$&\,\,$1$&\,\,$1$&-$1$&-$1$&\,\,$0$&\,\,$0$&\,\,$0$&$(1,1)$&-$2$&-$3$&\,\,$3$&\,\,$0$\\
&$+\beta\pm\gamma$&$(1,1,1)$&-$3$&\,\,$1$&-$1$&-$1$&\,\,$0$&\,\,$0$&\,\,$0$&$(\overline{3},1)$&\,\,$2$&\,\,$1$&\,\,$3$&\,\,$0$\\
&&$(3,1,1)$&-$1$&-$1$&\,\,$1$&\,\,$1$&\,\,$0$&\,\,$0$&\,\,$0$&$(1,1)$&\,\,$2$&\,\,$3$&-$3$&\,\,$0$\\
&&$(1,1,1)$&\,\,$3$&-$1$&\,\,$1$&\,\,$1$&\,\,$0$&\,\,$0$&\,\,$0$&$(3,1)$&-$2$&-$1$&-$3$&\,\,$0$\\
\hline
f&${S}+$&$(1,1,1)$&-$3$&\,\,$3$&\,\,$1$&-$1$&\,\,$0$&\,\,$0$&\,\,$0$&$(1,1)$&\,\,$2$&-$3$&\,\,$3$&\,\,$0$\\
&${b_{1}}+{b_{3}}$&$(1,1,1)$&-$3$&-$1$&-$3$&-$1$&\,\,$0$&\,\,$0$&\,\,$0$&$(1,1)$&\,\,$2$&-$3$&\,\,$3$&\,\,$0$\\
&$+\alpha\pm\gamma$&$(1,1,1)$&-$3$&-$1$&\,\,$1$&\,\,$3$&\,\,$0$&\,\,$0$&\,\,$0$&$(1,1)$&\,\,$2$&-$3$&\,\,$3$&\,\,$0$\\
&&$(1,1,1)$&\,\,$3$&\,\,$1$&-$1$&\,\,$1$&\,\,$0$&\,\,$0$&\,\,$0$&$(1,\overline{3})$&\,\,$2$&\,\,$3$&\,\,$1$&\,\,$0$\\
&&$(1,1,1)$&\,\,$3$&-$3$&-$1$&\,\,$1$&\,\,$0$&\,\,$0$&\,\,$0$&$(1,1)$&-$2$&\,\,$3$&-$3$&\,\,$0$\\
&&$(1,1,1)$&\,\,$3$&\,\,$1$&\,\,$3$&\,\,$1$&\,\,$0$&\,\,$0$&\,\,$0$&$(1,1)$&-$2$&\,\,$3$&-$3$&\,\,$0$\\
&&$(1,1,1)$&\,\,$3$&\,\,$1$&-$1$&-$3$&\,\,$0$&\,\,$0$&\,\,$0$&$(1,1)$&-$2$&\,\,$3$&-$3$&\,\,$0$\\
&&$(1,1,1)$&-$3$&-$1$&\,\,$1$&-$1$&\,\,$0$&\,\,$0$&\,\,$0$&$(1,3)$&-$2$&-$3$&-$1$&\,\,$0$\\
\hline
b&${b_{1}}+{b_{3}}$&$(\overline{3},1,1)$&\,\,$1$&-$1$&\,\,$1$&-$1$&\,\,$0$&\,\,$0$&\,\,$0$&$(1,1)$&\,\,$2$&-$3$&\,\,$3$&\,\,$0$\\
&$+\alpha\pm\gamma$&$(1,1,1)$&-$3$&-$1$&\,\,$1$&-$1$&\,\,$0$&\,\,$0$&\,\,$0$&$(\overline{3},1)$&-$2$&\,\,$1$&\,\,$3$&\,\,$0$\\
&&$(3,1,1)$&-$1$&\,\,$1$&-$1$&\,\,$1$&\,\,$0$&\,\,$0$&\,\,$0$&$(1,1)$&-$2$&\,\,$3$&-$3$&\,\,$0$\\
&&$(1,1,1)$&\,\,$3$&\,\,$1$&-$1$&\,\,$1$&\,\,$0$&\,\,$0$&\,\,$0$&$(3,1)$&\,\,$2$&-$1$&-$3$&\,\,$0$\\
\hline
\end{tabular}
\caption{Vector-like exotic states. All sectors, fermionic and bosonic, have CPT conjugates which are not displayed.}
\end{table}
\end{center}

\newpage
\begin{center}
\begin{table}[!ht]
\hspace*{-22 mm}
\begin{tabular}{|c|c|c|ccccccc|c|cccc|}
\hline
$F$&
SEC&$(C;L;R)$&$Q_{C}$&$Q_{\bar{\eta}^1}$&$Q_{\bar{\eta}^2}$&$Q_{\bar{\eta}^3}$&$Q_{\bar{y}^{3,6}}$&$Q_{\bar{y}^1\bar{w}^5}$&$Q_{\bar{w}^{2,4}}$&$SU(3)_{H_{1,2}}$&$Q_{\bar{\Phi}^1}$&$Q_{8}$&$Q_{9}$&$Q_{\bar{\Phi}^8}$\\
\hline
f&${S}+$&$(\overline{3},1,1)$&-$3$&-$1$&\,\,$1$&-$1$&\,\,$0$&\,\,$0$&\,\,$0$&$(1,1)$&\,\,$0$&\,\,$3$&-$3$&-$2$\\
&${1}+{b_{2}}$&$(1,1,1)$&\,\,$3$&-$1$&\,\,$1$&-$1$&\,\,$0$&\,\,$0$&\,\,$0$&$(1,\overline{3})$&\,\,$0$&\,\,$3$&\,\,$3$&\,\,$2$\\
&$+\alpha\pm\gamma$&$(3,1,1)$&\,\,$3$&\,\,$1$&-$1$&\,\,$1$&\,\,$0$&\,\,$0$&\,\,$0$&$(1,1)$&$\,\,0$&-$3$&$3$&\,\,$2$\\
&&$(1,1,1)$&-$3$&\,\,$1$&-$1$&\,\,$1$&\,\,$0$&\,\,$0$&\,\,$0$&$(1,3)$&\,\,$0$&-$3$&-$3$&-$2$\\
\hline
b&${1}+{b_{2}}$&$(1,1,1)$&-$3$&\,\,$3$&\,\,$1$&-$1$&\,\,$0$&\,\,$0$&\,\,$0$&$(1,1)$&\,\,$0$&\,\,$3$&-$3$&-$2$\\
&$+\alpha\pm\gamma$&$(1,1,1)$&-$3$&-$1$&-$3$&-$1$&\,\,$0$&\,\,$0$&\,\,$0$&$(1,1)$&\,\,$0$&\,\,$3$&-$3$&-$2$\\
&&$(1,1,1)$&-$3$&-$1$&\,\,$1$&\,\,$3$&\,\,$0$&\,\,$0$&\,\,$0$&$(1,1)$&\,\,$0$&\,\,$3$&-$3$&-$2$\\
&&$(1,1,1)$&-$3$&-$1$&\,\,$1$&-$1$&\,\,$0$&\,\,$0$&\,\,$0$&$(\overline{3},1)$&\,\,$0$&-$3$&-$3$&\,\,$2$\\
&&$(1,1,1)$&\,\,$3$&-$3$&-$1$&\,\,$1$&\,\,$0$&\,\,$0$&\,\,$0$&$(1,1)$&\,\,$0$&-$3$&\,\,$3$&\,\,$2$\\
&&$(1,1,1)$&\,\,$3$&\,\,$1$&\,\,$3$&\,\,$1$&\,\,$0$&\,\,$0$&\,\,$0$&$(1,1)$&\,\,$0$&-$3$&\,\,$3$&\,\,$2$\\
&&$(1,1,1)$&\,\,$3$&\,\,$1$&-$1$&-$3$&\,\,$0$&\,\,$0$&\,\,$0$&$(1,1)$&\,\,$0$&-$3$&\,\,$3$&\,\,$2$\\
&&$(1,1,1)$&\,\,$3$&\,\,$1$&-$1$&\,\,$1$&\,\,$0$&\,\,$0$&\,\,$0$&$(3,1)$&\,\,$0$&\,\,$3$&\,\,$3$&-$2$\\
\hline
f&${S}+$&$(1,1,1)$&-$3$&-$3$&-$1$&-$1$&\,\,$0$&\,\,$0$&\,\,$0$&$(1,1)$&\,\,$0$&\,\,$3$&-$3$&\,\,$2$\\
&${1}+{b_{1}}$&$(1,1,1)$&-$3$&\,\,$1$&\,\,$3$&-$1$&\,\,$0$&\,\,$0$&\,\,$0$&$(1,1)$&\,\,$0$&\,\,$3$&-$3$&\,\,$2$\\
&$+\beta\pm\gamma$&$(1,1,1)$&-$3$&\,\,$1$&-$1$&\,\,$3$&\,\,$0$&\,\,$0$&\,\,$0$&$(1,1)$&\,\,$0$&\,\,$3$&-$3$&\,\,$2$\\
&&$(1,1,1)$&-$3$&\,\,$1$&-$1$&-$1$&\,\,$0$&\,\,$0$&\,\,$0$&$(3,1)$&\,\,$0$&-$3$&-$3$&-$2$\\
&&$(1,1,1)$&\,\,$3$&\,\,$3$&\,\,$1$&\,\,$1$&\,\,$0$&\,\,$0$&\,\,$0$&$(1,1)$&\,\,$0$&-$3$&\,\,$3$&-$2$\\
&&$(1,1,1)$&\,\,$3$&-$1$&-$3$&\,\,$1$&\,\,$0$&\,\,$0$&\,\,$0$&$(1,1)$&\,\,$0$&-$3$&\,\,$3$&-$2$\\
&&$(1,1,1)$&\,\,$3$&-$1$&\,\,$1$&-$3$&\,\,$0$&\,\,$0$&\,\,$0$&$(1,1)$&\,\,$0$&-$3$&\,\,$3$&-$2$\\
&&$(1,1,1)$&\,\,$3$&-$1$&\,\,$1$&\,\,$1$&\,\,$0$&\,\,$0$&\,\,$0$&$(\overline{3},1)$&\,\,$0$&$3$&$3$&\,\,$2$\\
\hline
b&${1}+{b_{1}}$&$(3,1,1)$&\,\,$3$&\,\,$1$&-$1$&-$1$&\,\,$0$&\,\,$0$&\,\,$0$&$(1,1)$&\,\,$0$&\,\,$3$&-$3$&\,\,$2$\\
&$+\beta \pm
\gamma$&$(1,1,1)$&-$3$&\,\,$1$&-$1$&-$1$&\,\,$0$&\,\,$0$&\,\,$0$&$(1,3)$&\,\,$0$&\,\,$3$&\,\,$3$&-$2$\\
&&$(\overline{3},1,1)$&-$3$&-$1$&\,\,$1$&\,\,$1$&\,\,$0$&\,\,$0$&\,\,$0$&$(1,1)$&\,\,$0$&-$3$&\,\,$3$&-$2$\\
&&$(1,1,1)$&\,\,$3$&-$1$&\,\,$1$&\,\,$1$&\,\,$0$&\,\,$0$&\,\,$0$&$(1,\overline{3})$&\,\,$0$&-$3$&-$3$&\,\,$2$\\
\hline
\end{tabular}
\caption{Vector-like exotic states. All sectors, fermionic and bosonic, have CPT conjugates which are not displayed.}
\end{table}
\end{center}

\newpage
\begin{center}
\begin{table}[!ht]
\hspace*{-22 mm}
\begin{tabular}{|c|c|c|ccccccc|c|cccc|}
\hline
$F$&
SEC&$(C;L;R)$&$Q_{C}$&$Q_{\bar{\eta}^1}$&$Q_{\bar{\eta}^2}$&$Q_{\bar{\eta}^3}$&$Q_{\bar{y}^{3,6}}$&$Q_{\bar{y}^1\bar{w}^5}$&$Q_{\bar{w}^{2,4}}$&$SU(3)_{H_{1,2}}$&$Q_{\bar{\Phi}^1}$&$Q_{8}$&$Q_{9}$&$Q_{\bar{\Phi}^8}$\\
\hline
f&${1}+{b_{2}}$&$(1,2,1)$&\,\,$0$&\,\,$0$&-$2$&-$2$&\,\,$2$&\,\,$0$&\,\,$0$&$(1,1)$&-$2$&\,\,$0$&\,\,$0$&\,\,$2$\\
&$+{{b_{3}}}+2\gamma$&$(1,2,1)$&\,\,$0$&\,\,$0$&-$2$&-$2$&-$2$&\,\,$0$&\,\,$0$&$(1,1)$&\,\,$2$&\,\,$0$&\,\,$0$&-$2$\\
&&$(1,1,2)$&\,\,$0$&\,\,$0$&\,\,$2$&\,\,$2$&\,\,$2$&\,\,$0$&\,\,$0$&$(1,1)$&\,\,$2$&\,\,$0$&\,\,$0$&-$2$\\
&&$(1,1,2)$&\,\,$0$&\,\,$0$&\,\,$2$&\,\,$2$&-$2$&\,\,$0$&\,\,$0$&$(1,1)$&-$2$&\,\,$0$&\,\,$0$&\,\,$2$\\
\hline
b&${S}+$&$(1,2,1)$&\,\,$0$&\,\,$0$&\,\,$2$&\,\,$2$&-$2$&\,\,$0$&\,\,$0$&$(1,1)$&-$2$&\,\,$0$&\,\,$0$&\,\,$2$\\
&${1}+{b_{2}}$&$(1,2,1)$&\,\,$0$&\,\,$0$&\,\,$2$&\,\,$2$&\,\,$2$&\,\,$0$&\,\,$0$&$(1,1)$&\,\,$2$&\,\,$0$&\,\,$0$&-$2$\\
&$+{b_{3}}+2\gamma$&$(1,1,2)$&\,\,$0$&\,\,$0$&-$2$&-$2$&-$2$&\,\,$0$&\,\,$0$&$(1,1)$&\,\,$2$&\,\,$0$&\,\,$0$&-$2$\\
&&$(1,1,2)$&\,\,$0$&\,\,$0$&-$2$&-$2$&\,\,$2$&\,\,$0$&\,\,$0$&$(1,1)$&-$2$&\,\,$0$&\,\,$0$&\,\,$2$\\
\hline
f&${1}+{b_{1}}$&$(1,2,1)$&\,\,$0$&-$2$&\,\,$0$&-$2$&\,\,$0$&\,\,$2$&\,\,$0$&$(1,1)$&-$2$&$0$&$0$&\,\,$2$\\
&$+{{b_{3}}}+2\gamma$&$(1,2,1)$&\,\,$0$&-$2$&\,\,$0$&-$2$&\,\,$0$&-$2$&\,\,$0$&$(1,1)$&\,\,$2$&$0$&$0$&-$2$\\
&&$(1,1,2)$&\,\,$0$&\,\,$2$&\,\,$0$&\,\,$2$&\,\,$0$&\,\,$2$&\,\,$0$&$(1,1)$&\,\,$2$&$0$&$0$&-$2$\\
&&$(1,1,2)$&\,\,$0$&\,\,$2$&\,\,$0$&\,\,$2$&\,\,$0$&-$2$&\,\,$0$&$(1,1)$&-$2$&$0$&$0$&\,\,$2$\\
\hline
b&${S}+$&$(1,2,1)$&\,\,$0$&\,\,$2$&\,\,$0$&\,\,$2$&\,\,$0$&-$2$&\,\,$0$&$(1,1)$&-$2$&$0$&$0$&\,\,$2$\\
&${1}+{b_{1}}$&$(1,2,1)$&\,\,$0$&\,\,$2$&\,\,$0$&\,\,$2$&\,\,$0$&\,\,$2$&\,\,$0$&$(1,1)$&\,\,$2$&$0$&$0$&-$2$\\
&$+{b_{3}}+2\gamma$&$(1,1,2)$&\,\,$0$&-$2$&\,\,$0$&-$2$&\,\,$0$&-$2$&\,\,$0$&$(1,1)$&\,\,$2$&$0$&$0$&-$2$\\
&&$(1,1,2)$&\,\,$0$&-$2$&\,\,$0$&-$2$&\,\,$0$&\,\,$2$&\,\,$0$&$(1,1)$&-$2$&$0$&$0$&\,\,$2$\\
\hline
f&${1}+{b_{1}}$&$(1,2,1)$&\,\,$0$&-$2$&-$2$&\,\,$0$&\,\,$0$&\,\,$0$&\,\,$2$&$(1,1)$&-$2$&$0$&$0$&\,\,$2$\\
&$+{{b_{2}}}+2\gamma$&$(1,2,1)$&\,\,$0$&-$2$&-$2$&\,\,$0$&\,\,$0$&\,\,$0$&-$2$&$(1,1)$&\,\,$2$&$0$&$0$&-$2$\\
&&$(1,1,2)$&\,\,$0$&\,\,$2$&\,\,$2$&\,\,$0$&\,\,$0$&\,\,$0$&\,\,$2$&$(1,1)$&\,\,$2$&$0$&$0$&-$2$\\
&&$(1,1,2)$&\,\,$0$&\,\,$2$&\,\,$2$&\,\,$0$&\,\,$0$&\,\,$0$&-$2$&$(1,1)$&-$2$&$0$&$0$&\,\,$2$\\
\hline
b&${S}+$&$(1,2,1)$&\,\,$0$&\,\,$2$&\,\,$2$&\,\,$0$&\,\,$0$&\,\,$0$&-$2$&$(1,1)$&-$2$&$0$&$0$&\,\,$2$\\
&${1}+{b_{1}}$&$(1,2,1)$&\,\,$0$&\,\,$2$&\,\,$2$&\,\,$0$&\,\,$0$&\,\,$0$&\,\,$2$&$(1,1)$&\,\,$2$&$0$&$0$&-$2$\\
&$+{b_{2}}+2\gamma$&$(1,1,2)$&\,\,$0$&-$2$&-$2$&\,\,$0$&\,\,$0$&\,\,$0$&-$2$&$(1,1)$&\,\,$2$&$0$&$0$&-$2$\\
&&$(1,1,2)$&\,\,$0$&-$2$&-$2$&\,\,$0$&\,\,$0$&\,\,$0$&\,\,$2$&$(1,1)$&-$2$&$0$&$0$&\,\,$2$\\
\hline
f&${S}+$&$(1,1,1)$&-$6$&\,\,$0$&\,\,$0$&-$2$&\,\,$0$&\,\,$0$&\,\,$0$&$(1,1)$&\,\,$2$&\,\,$0$&\,\,$0$&\,\,$2$\\
&${1}+{b_{3}}$&$(\overline{3},1,1)$&-$2$&\,\,$0$&\,\,$0$&\,\,$2$&\,\,$0$&\,\,$0$&\,\,$0$&$(1,1)$&-$2$&\,\,$0$&\,\,$0$&-$2$\\
&$+\alpha+\beta$&$(1,1,1)$&\,\,$6$&\,\,$0$&\,\,$0$&\,\,$2$&\,\,$0$&\,\,$0$&\,\,$0$&$(1,1)$&-$2$&\,\,$0$&\,\,$0$&-$2$\\
&$+2\gamma$&$(3,1,1)$&\,\,$2$&\,\,$0$&\,\,$0$&-$2$&\,\,$0$&\,\,$0$&\,\,$0$&$(1,1)$&\,\,$2$&\,\,$0$&\,\,$0$&\,\,$2$\\
\hline

b&${1}+{b_{3}}$&$(1,1,1)$&\,\,$6$&\,\,$0$&\,\,$0$&\,\,$2$&\,\,$0$&\,\,$0$&\,\,$0$&$(1,1)$&-$2$&\,\,$0$&\,\,$0$&-$2$\\
&$+\alpha+\beta$&$(1,1,1)$&-$6$&\,\,$0$&\,\,$0$&-$2$&\,\,$0$&\,\,$0$&\,\,$0$&$(1,1)$&\,\,$2$&\,\,$0$&\,\,$0$&\,\,$2$\\
&$+2\gamma$&$(3,1,1)$&\,\,$2$&\,\,$0$&\,\,$0$&-$2$&\,\,$0$&\,\,$0$&\,\,$0$&$(1,1)$&\,\,$2$&\,\,$0$&\,\,$0$&\,\,$2$\\
&&$(\overline{3},1,1)$&-$2$&\,\,$0$&\,\,$0$&\,\,$2$&\,\,$0$&\,\,$0$&\,\,$0$&$(1,1)$&-$2$&\,\,$0$&\,\,$0$&-$2$\\
\hline
\end{tabular}
\caption{Vector-like exotic states. All sectors, fermionic and bosonic, have CPT conjugates which are not displayed.}
\end{table}
\end{center}

\newpage


\bigskip
\medskip

\bibliographystyle{unsrt}

\end{document}